\documentclass[12pt]{article}

\usepackage[top=1in,bottom=1in,left=1in,right=1in]{geometry} 
\usepackage{array}     
\newcolumntype{L}[1]{>{\raggedright\arraybackslash}p{#1}} 
\usepackage{xltabular}
\usepackage{booktabs}
\usepackage[doublespacing]{setspace}
\usepackage{amsfonts, amssymb, amsmath, amsthm}
\usepackage{graphicx, hyperref, caption, subcaption, xcolor, verbatim, titlesec, booktabs, multirow, enumitem, tocloft, bbm, mathrsfs}
\usepackage{mathtools,pdflscape}
\usepackage[sans]{dsfont}
\usepackage[scr=rsfs,cal=boondox]{mathalpha}
\hypersetup{pdfborder = {0 0 0},colorlinks=true,allcolors=blue}
\usepackage{natbib}
\usepackage[title]{appendix}
\usepackage{placeins}
\usepackage{threeparttable}
\usepackage[normalem]{ulem}

\allowdisplaybreaks[1]

\DeclareMathOperator*{\argmin}{arg\,min}
\DeclareMathOperator{\T}{T}
\DeclareMathOperator{\CI}{I}

\newcommand{\reg}{\mathtt{reg}}

\renewcommand{\P}{\mathbb{P}}
\renewcommand{\d}{\mathcal{d}}
\newcommand{\E}{\mathbb{E}}

\newcommand{\Indicator}{\mathds{1}}

\newcommand{\Haus}{\mathfrak{H}}
\newcommand{\Jacob}{\operatorname{J}\hspace{-.025in}}

\newcommand{\bb}{\mathbf{b}}

\newcommand{\br}{\mathbf{r}}

\newcommand{\bu}{\mathbf{u}}

\newcommand{\bx}{\mathbf{x}}
\newcommand{\bX}{\mathbf{X}}

\newcommand{\bZ}{\mathbf{Z}}

\newcommand{\bbeta}{\boldsymbol{\beta}}

\newcommand{\biota}{\boldsymbol{\iota}}

\newcommand{\A}{\mathcal{A}}
\newcommand{\B}{\mathcal{B}}

\newcommand{\X}{\mathcal{X}}

\begin{document}
 
\title{Boundary Discontinuity Designs: Theory and Practice\thanks{This chapter was prepared for the invited session on ``Causal Inference and Statistical Decisions'' at the 2025 World Congress of the Econometric Society, Seoul (Korea). We thank Alberto Abadie, Joshua Angrist, Sandra Black, David Card, Xiaohong Chen, Boris Hanin, Leander Heldring, Thomas Holmes, Kosuke Imai, Rafael Lalive, Oliver Linton, Xinwei Ma, Francesca Molinari, Jesse Shapiro, J\"org Stoye, Jeff Wooldridge, and Joseph Zweimuller for comments and discussions. Cattaneo and Titiunik gratefully acknowledge financial support from the National Science Foundation (SES-2019432, DMS-2210561, SES-2241575 and SES-2342226). Cattaneo gratefully acknowledges financial support from the National Institute for Food and Agriculture (NIFA) through grant 2024-67023-42704, and the Data-Driven Social Science initiative at Princeton University.}\bigskip}

\author{Matias D. Cattaneo\thanks{Department of Operations Research and Financial Engineering, Princeton University.} \and
	    Roc\'{i}o Titiunik\thanks{Department of Politics, Princeton University.} \and
	    Ruiqi (Rae) Yu\thanks{Department of Operations Research and Financial Engineering, Princeton University.} 
	    }
\maketitle

\begin{abstract}
    The boundary discontinuity (BD) design is a non-experimental method for identifying causal effects that exploits a thresholding rule based on a bivariate score and a boundary curve. This widely used method generalizes the univariate regression discontinuity design but introduces unique challenges arising from its multidimensional nature. We synthesize over $80$ empirical papers that use the BD design, tracing the method's application from its formative stages to its implementation in modern research. We also overview ongoing theoretical and methodological research on identification, estimation, and inference for BD designs employing local polynomial regression, and offer recommendations for practice.
\end{abstract}

\thispagestyle{empty}

\clearpage
\tableofcontents
\thispagestyle{empty}

\clearpage
\setcounter{page}{1}

\section{Introduction}\label{sec: Introduction}

The Regression Discontinuity (RD) design is one of the leading observational methods for program evaluation and causal inference \citep[see][for overviews and further references]{Abadie-Cattaneo_2018_ARE,Hernan-Robins_2020_Book}. In its canonical form, a binary treatment is assigned to units whose value of a univariate score is equal to or above a known scalar cutoff, and not assigned to units whose score value is below the cutoff. Under the assumption that all observable and unobservable pretreatment characteristics vary smoothly at the cutoff, the discontinuous change generated by the hard-thresholding treatment assignment rule can be used to learn about causal treatment effects. \cite{Lee-Lemieux_2010_JEL} offer an early review and list of RD empirical applications, \cite{Cattaneo-Titiunik_2022_ARE} give a recent review of the RD methodological literature, and \cite{Cattaneo-Idrobo-Titiunik_2020_CUP,Cattaneo-Idrobo-Titiunik_2024_CUP} provide a practical introduction to modern RD methods.

The Boundary Discontinuity (BD) design generalizes the RD design: the score has two dimensions instead of one, and the treatment is assigned to units according to the location of their score relative to a known boundary curve that splits the support into two disjoint regions. This setup is sometimes called the Multi-Score RD design \citep{Papay-Willett-Murnane_2011_JoE,Reardon-Robinson_2012_JREE,Wong-Steiner-Cook_2013_JEBS}, or the Geographic RD design \citep{Keele-Titiunik_2015_PA,Keele-Titiunik-Zubizarreta_2015_JRSSA,Keele-Titiunik_2016_PSRM,Keele-etal_2017_AIE,Galiani-McEwan-Quistorff_2017_AIE,Rischard-Branson-Miratrix-Bornn_2021_JASA,Diaz-Zubizarreta_2023_AOAS}. We discuss the empirical strategies most commonly employed for the analysis and interpretation of BD designs, and review ongoing theoretical and methodological results characterizing their econometric properties \citep{Cattaneo-Titiunik-Yu_2025_BDD-Distance,Cattaneo-Titiunik-Yu_2025_BDD-Location,Cattaneo-Titiunik-Yu_2025_BDD-Pooling}.

We start in Section \ref{sec: The Boundary Discontinuity Design} by introducing the BD design and presenting a review of the empirical literature in Economics, Political Science, Education, and other disciplines. We report over $80$ empirical papers in Table \ref{tab:LitReview}, which form the basis of the rest of our discussion. We find that the overwhelming majority of papers collapse the bivariate score into a univariate distance score measuring the shortest distance from the unit's location to the treatment assignment boundary, and then report a single average treatment effect estimated by pooling all the observations that are close to the boundary, regardless of their specific location. We refer to this empirical strategy as \textit{pooling-based methods}. In contrast, only a handful of empirical papers have investigated heterogeneity by localizing to specific regions or points on the boundary, despite the usefulness of this rich information for policy evaluation and decision-making.

In Section \ref{sec: Pooling-Based Methods}, we discuss pooling-based methods. We begin with the seminal work of \cite{Card-Krueger_1994_AER}. Although their paper is often cited as a canonical example of a difference-in-differences design, these authors are one of the first to argue that reducing geographic distance may increase comparability among units. We then discuss \cite{holmes1998effect} and \cite{black1999better}, who are among the first to employ localization to the assignment boundary explicitly as the central empirical strategy. Subsequent empirical work has recognized the importance of localization to the boundary, and has considered different empirical strategies for treatment effect estimation via local flexible regression methods. One of the most influential papers is \cite{dell2010persistent}, who incorporated a polynomial expansion of the bivariate location score as part of the local regression specification. More recently, an alternative empirical approach views the estimation based on distance to the boundary as a pooled univariate RD design, and thus estimates treatment effects using a local regression including a polynomial expansion of the univariate distance score---recent examples include \cite{ito2020willingness} and \cite{dehdari2022origins}. Many empirical papers also include boundary-segment fixed effects in their local regression specifications, and some include interactions with the treatment indicator and/or between the polynomial expansions of the univariate distance score and the bivariate location score. Furthermore, either to simplify the problem or because distinct treatments occur in different regions of the boundary, some papers reduce the BD design to boundary-segment-specific univariate RD designs---see \cite{ou2010leave}, \cite{LondonoVelezRodriguezSanchez_2020_AEJ} and \cite{salti2022impact}. We synthesize these various empirical strategies into eight distinct pooled local regression specifications.

While pooling-based methods are widely used in empirical work, their econometric properties are not well understood. In ongoing research, \cite{Cattaneo-Titiunik-Yu_2025_BDD-Pooling} formally study identification, estimation, and inference for these methods, recast as a univariate RD design based on the shortest distance to the assignment boundary. The results leverage geometric measure theory \citep{federer2014geometric} upon recognizing the assignment boundary as a one-dimensional submanifold curve on the plane, and enable estimation and inference using state-of-the-art RD methods \citep{Calonico-Cattaneo-Titiunik_2014_ECMA,Calonico-Cattaneo-Farrell-Titiunik_2019_RESTAT,Calonico-Cattaneo-Farrell_2020_ECTJ}. Some of this work is conceptually related to a recent paper by \cite{chen2025semiparametric}, who study estimation and inference for integral functionals on submanifolds when using nonparametric sieve estimation.

The boundary average treatment effect parameter emerging from the pooling-based methods is useful, but the richness of the two-dimensional BD assignment offers the opportunity to explore heterogeneous treatment effects along the assignment boundary. In geographic applications, the heterogeneity is directly linked to the geographic location of the units, providing useful information about how treatment effects vary in space; in non-geographic settings, the heterogeneity is directly linked to unit features that are captured by the bivariate score, and hence its interpretation is necessarily application-specific.

In Section \ref{sec: Distance-based and Location-Based Methods}, we discuss heterogeneity analysis and subsequent aggregation of causal treatment effects in BD designs using the boundary average treatment effect curve (BATEC), which captures the average treatment effect for each point along the assignment boundary. We also summarize the results in \cite{Cattaneo-Titiunik-Yu_2025_BDD-Distance} and \cite{Cattaneo-Titiunik-Yu_2025_BDD-Location}. These papers study identification, estimation, and inference for BATEC and transformations thereof, from two distinct perspectives: \cite{Cattaneo-Titiunik-Yu_2025_BDD-Distance} studies boundary-point-specific distance-based methods, which are directly motivated by the pooling-based approaches discussed in Section \ref{sec: Pooling-Based Methods}, while \cite{Cattaneo-Titiunik-Yu_2025_BDD-Location} studies methods based on the untransformed bivariate location. For each approach, the local polynomial estimation and inference results are valid both pointwise and uniformly along the assignment boundary, in addition to providing the basis for constructing other treatment effect estimators.

We offer recommendations for practice and concluding remarks in Section \ref{sec: Recommendations for Practice}. Our recommended methods can be implemented using the general-purpose software packages \texttt{rdrobust}, \texttt{rdhte}, \texttt{rdmulti}, and \texttt{rd2d}, located at \url{https://rdpackages.github.io/}, where readers can also find replication files and related references.

\section{The Boundary Discontinuity Design}\label{sec: The Boundary Discontinuity Design}

Each unit $i$ in the study, $i=1,\ldots, n$, has a continuous bivariate score $\bX_i=(X_{1i},X_{2i})$ that takes values in the set $\X\subseteq\mathbb{R}^2$. The assignment of units to the treatment or control condition depends on the location of their score $\bX_i$ relative to a known one-dimensional boundary curve $\B$ that splits $\X$ into two disjoint regions: $\X = \A_0 \cup \A_1$, with $\A_0$ and $\A_1$ the disjoint (connected) control and treatment regions, respectively, and $\B = \mathtt{bd}(\A_0) \cap \mathtt{bd}(\A_1)$, where $\mathtt{bd}(\A_t)$ denotes the boundary of the set $\A_t$. We assume that the boundary belongs to the treatment group, that is, $\mathtt{bd}(\A_1)\subset\A_1$ and $\B\cap\A_0=\emptyset$. The observed outcome variable is $Y_i$.

An ancestor of the BD design  was used by  \cite{Card-Krueger_1994_AER} in their seminal study of the effects of the minimum wage on employment. In their analysis, the treatment of interest is the increase of the state minimum wage in New Jersey adopted on April 1, 1992, which rose the minimum wage to \$5.05. The authors compared a sample of fast food restaurants in New Jersey to a sample of fast food restaurants in eastern Pennsylvania, a state that shares a border with New Jersey and where the minimum wage did not increase. After the increase in New Jersey, the average starting wage was \$5.08 in the New Jersey sample and \$4.62 in the Pennsylvania sample.

The rationale for comparing restaurants in New Jersey to restaurants in eastern Pennsylvania was that, by focusing on close geographic areas, these restaurants would be subject to similar economic conditions that would otherwise confound the effects of changes to the minimum wage. Although \citeauthor{Card-Krueger_1994_AER}'s main research strategy was a difference-in-differences strategy comparing NJ and PA before and after the increase, they explicitly mention the close geographic proximity between the areas to justify their choice of the comparison group, as they believed that fast-food stores in eastern Pennsylvania formed a ``natural basis for comparison with the experiences of restaurants in New Jersey'' due to Pennsylvania being a ``nearby state'' \citep[][p. 773]{Card-Krueger_1994_AER}.

The use of close proximity as the basis for reducing potential confounders and enhancing the credibility of causal interpretations bears a direct connection to the BD design. The main difference is that \cite{Card-Krueger_1994_AER} included non-adjacent areas in New Jersey and Pennsylvania (e.g., the New Jersey sample included restaurants in the New Jersey shore, far away from the Pennsylvania border), and did not directly use the unit's distance to the boundary in their analysis. The explicit use of distance to the assignment boundary is central to the BD design, which is why we consider \cite{Card-Krueger_1994_AER} to be an ancestor rather than an instance of the BD design itself: while the authors leverage the change in treatment assignment induced by a geographic boundary, they did not fully localize to the boundary itself in their analysis. (See \citet{dube2010minimum} for an example of a minimum wage study in the spirit of \cite{Card-Krueger_1994_AER} that uses distance explicitly in the analysis.)


\citet{holmes1998effect} and \cite{black1999better} are the earliest empirical examples employing the BD design that we could find. \citet{holmes1998effect} used a BD design to study the effect of pro-business policies on manufacturing, comparing states that adopted right-to-work laws with adjacent states that did not. Using the longitude and latitude coordinates of each county's centroid, he calculated the minimum distance of the centroid to the assignment border, and focused the analysis on counties close to the border as a way to ``control for differences across states in these various characteristics that are unrelated to policy'' \citep[][p. 671]{holmes1998effect}. Similarly, \cite{black1999better} used a BD design to infer the quality of public schools from housing prices. She compared house prices on opposite sides of school attendance district boundaries, restricting the sample to houses close to the boundary. The justification for this localization was that houses on opposite sides of the boundary within a small area around it would be similar in all characteristics except for school quality---as the latter changes discontinuously at the attendance district boundary. In these examples, the bivariate score is the pair of latitude and longitude coordinates that determines the location of each unit (county or house), and the assignment boundary is the geographic border between two adjacent states or school attendance districts; the treatment is the policy or feature that changes abruptly at the border---labor policy for \citet{holmes1998effect} and school quality for \citet{black1999better}. This type of BD design is often referred to as a Geographic RD design.

One of the most common non-geographic applications of the BD design occurs in education when a treatment is given on the basis of two exam scores. For example, \cite{ou2010leave} studied the effect of a high school exit exam on  the likelihood of dropping out of high school. The exam had a mathematics and a language arts component, each of which was graded separately; students must achieve a minimum proficiency in each section in order to pass the exam. The bivariate score in this case is the pair of mathematics and language test scores, and the treatment is passing the exam.

\begin{figure}
    \centering
    \begin{subfigure}[b]{0.45\textwidth}
        \centering
        \includegraphics[width=\linewidth]{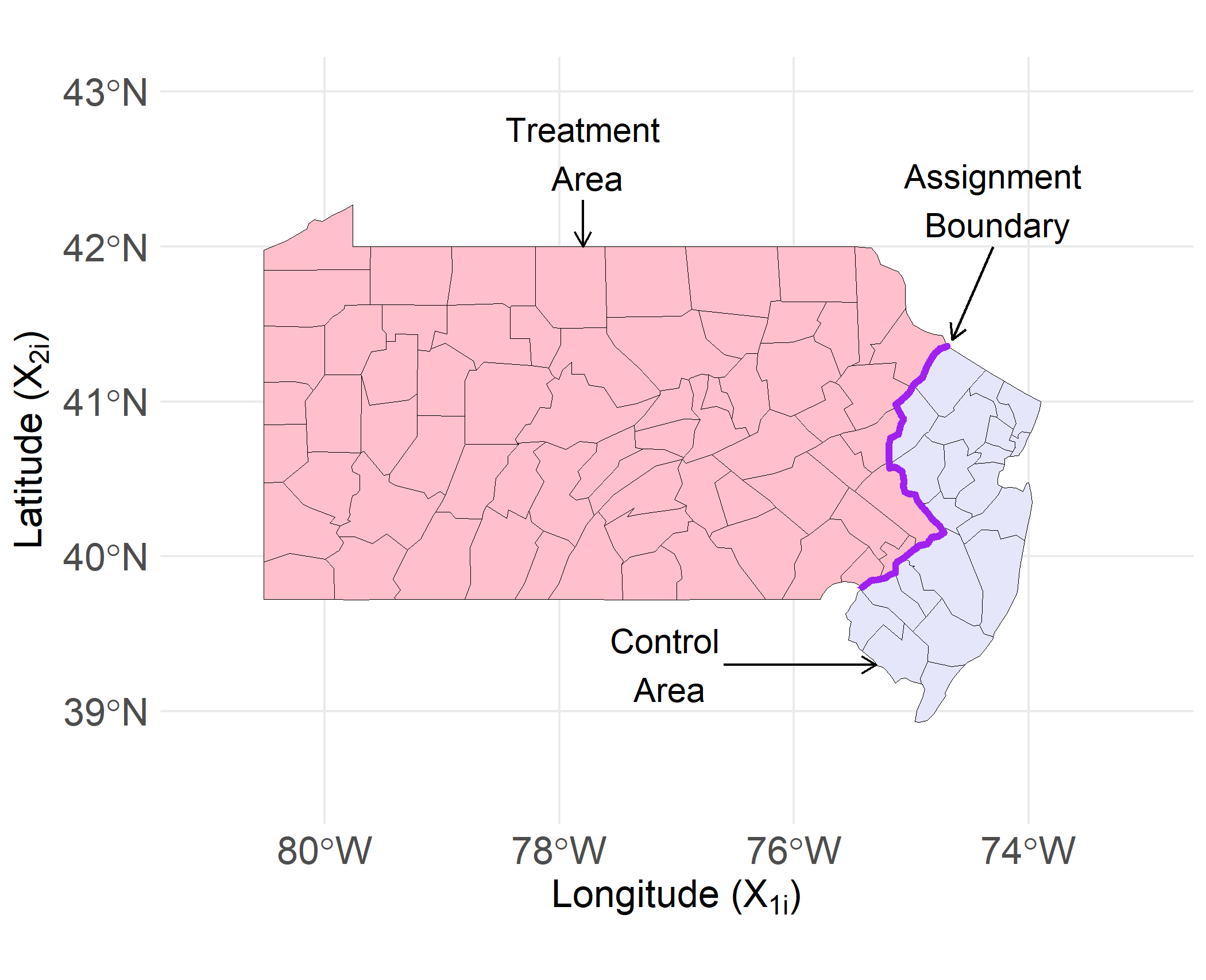}
        \caption{Geographic BD Design}
        \label{fig:BDgeo}
    \end{subfigure}
    \hspace{.5in}
    \begin{subfigure}[b]{0.45\textwidth}
        \centering
        \includegraphics[width=\linewidth]{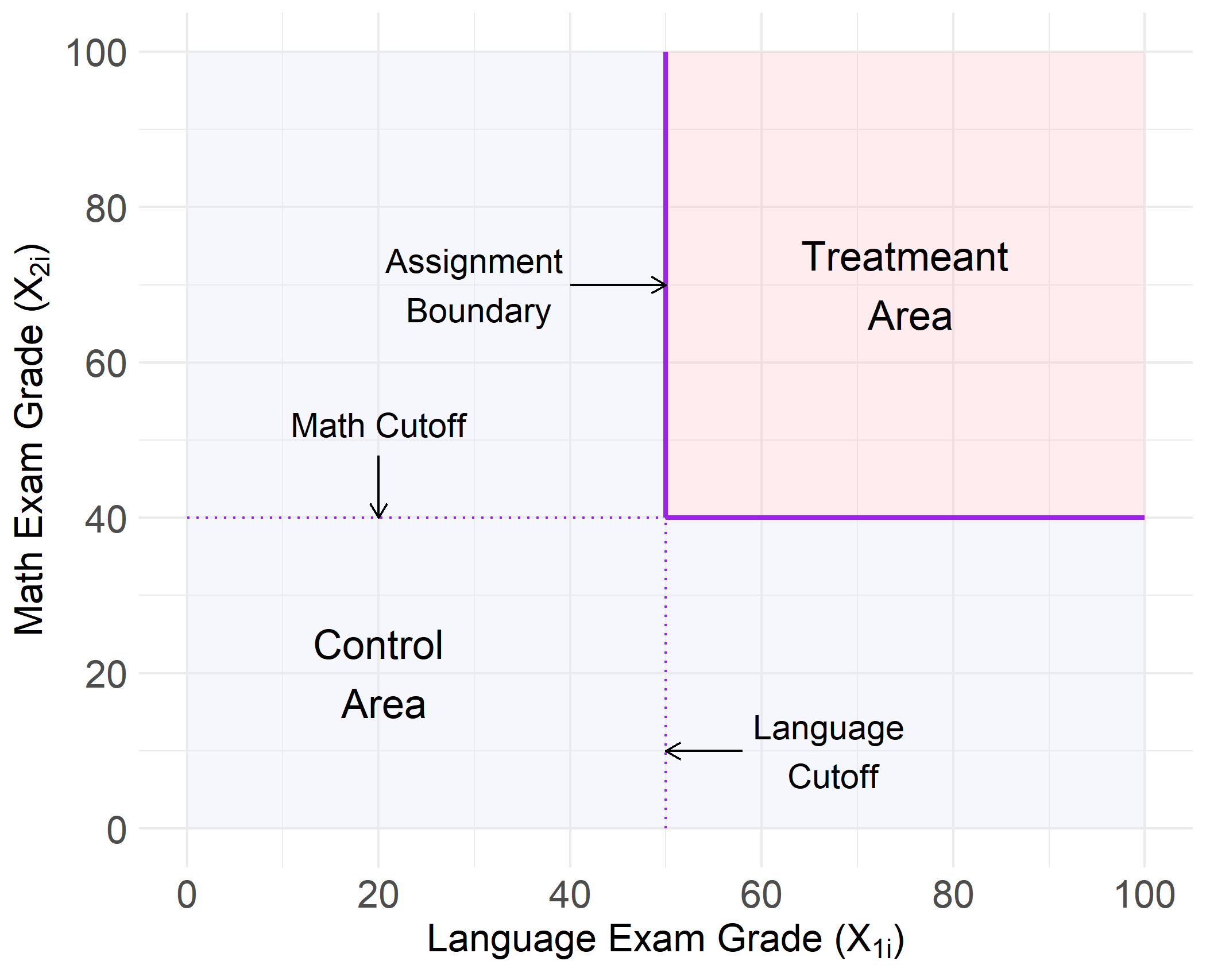}
        \caption{Non-geographic BD Design}
        \label{fig:BDnongeo}
    \end{subfigure}    
    \caption{BD Design Illustration}
    \label{fig:BDillustration}
\end{figure}

Figure \ref{fig:BDillustration} illustrates the boundary and treatment assignment regions in two stylized settings corresponding to the geographic and non-geographic examples. Figure \ref{fig:BDgeo} illustrates the geographic BD design case, where units are split into adjacent treated and control areas according to their location with respect to a geographic border, as in \cite{black1999better} and \citet{holmes1998effect}. (This figure shows the New Jersey-Pennsylvania border.) Figure \ref{fig:BDnongeo} illustrates the non-geographic setting, using as example the study by \cite{ou2010leave} where students receive two exam scores and a treatment is given only to those individuals who score above a minimum cutoff in each exam. In geographic applications, the bivariate score is always composed of the geographic coordinates of the units. In non-geographic applications, the bivariate score could include different types of variables; in addition to test scores from multiple exams, examples include shares of airline passengers in origin and destination cities \citep{snider2015barriers}, tax rates and income ratios \citep{egger2015impact}, systolic and diastolic blood pressure measurements \citep{dai2022effects}, different component scores in a means-testing formula \citep{salti2022impact,kampfen2024heterogeneous}, and population counts at different time  periods \citep{hinnerich2014democracy}.

Figure \ref{fig:BDillustration} illustrates an important distinction between the geographic and non-geographic examples. In non-geographic examples, the function that describes the boundary tends to be known and is typically linear with few irregularities and few kink points. In contrast, in geographic BD designs the boundary is typically the border between political or administrative subdivisions; as a consequence, its shape is only available as coordinates on a map, and its geometry is complex: in the best case, it will be piecewise linear with kink points, but it will be more irregular in many applications. In his early study, \citet[][pp. 681--682]{holmes1998effect} recognized this challenge and discussed explicitly the complex boundary shapes created by geographic borders. Indeed, since the influential work by \cite{mandelbrot1967long,mandelbrot1983fractal}, there has been an ongoing debate among mathematicians and philosophers about whether geographic borders (and other shapes in nature) are fractals \citep[see][and references therein]{avnir1998geometry}. From a methodological perspective, as recently formalized in \cite{Cattaneo-Titiunik-Yu_2025_BDD-Distance,Cattaneo-Titiunik-Yu_2025_BDD-Location,Cattaneo-Titiunik-Yu_2025_BDD-Pooling}, the geometry of the assignment boundary in BD designs has fundamental implications for identification, estimation, and inference. We will review some of these recent findings in the upcoming sections.

In the standard RD design, the distance between the univariate score and the single cutoff is naturally measured by the Euclidean distance, and observations are close to the cutoff $c$ when their univariate score $X_i$ implies a small distance $|X_i - c|$. In contrast, in the BD design, closeness to the boundary can be defined in different ways, depending on whether the parameter of interest is defined relative to a specific point on the boundary, or as an aggregate thereof along the entire boundary, such as an average or a supremum. Furthermore, due to its multivariate nature, the analysis can depend on the specific notion of distance used. 

In the upcoming sections, we discuss the two main approaches to the analysis of BD designs. The first approach uses the variation in treatment assignment induced by the entire boundary simultaneously, and thus pools all observations sufficiently close to the boundary in a single analysis, producing a single average treatment effect across all boundary points. This pooling approach relies on localization to $\B$ by first computing the distance between each observation's location and the nearest point on the boundary, denoted by $D_i$, and then retaining only those observations for which this distance is no larger than a specific bandwidth, denoted by $h$. The advantage of this approach is that it naturally mimics the univariate RD design, using the bivariate location of each unit to build a univariate score that is used directly in a standard univariate RD analysis. This approach produces a single average treatment effect estimate.

The second approach focuses on estimating the average treatment effect at each point on the boundary, performing localization relative to the specific point of analysis, thereby retaining observations whose distance to that point is no greater than a chosen bandwidth. Estimation and inference are conducted separately at each boundary point $\bx\in\B$, capturing the heterogeneity of treatment effects along the boundary. Aggregate causal effects can be recovered by averaging (or applying other transformations to) these boundary-point average treatment effects, including as a special case the average treatment effect along the entire boundary that is the focus of the pooling approach. The boundary-point approach can be seen as focusing on heterogeneous average treatment effects that are then used as building blocks to construct more aggregate causal parameters. The advantage of this approach is that it captures the full richness of the BD design by first learning the heterogeneity of the average treatment effect along the boundary, which usually summarizes relevant unit characteristics such as geographic location, academic performance, or economic need.

\subsection{BD Designs in Practice}

We searched the academic literature in multiple social sciences to learn how applied researchers implemented the BD design. We conducted our search in four general queries. In our initial Economics query, we included ten leading journals in Economics---\textit{Journal of Political Economy}, \textit{Review of Economic Studies}, \textit{Econometrica}, \textit{American Economic Review}, \textit{Quarterly Journal of Economics}, \textit{Review of Economics and Statistics}, \textit{American Economic Journal: Applied Economics}, \textit{American Economic Journal: Economic Policy}, \textit{American Economic Journal: Macroeconomics}, and \textit{American Economic Journal: Microeconomics}. In our initial Political Science query, we searched six leading Political Science journals---\textit{American Political Science Review}, \textit{American Journal of Political Science}, \textit{Journal of Politics}, \textit{Political Science Research and Methods}, \textit{British Journal of Political Science}, and \textit{Quarterly Journal of Political Science}. In both cases, we searched for terms like ``boundary discontinuity'', ``multivariate regression discontinuity'', and ``multiscore regression discontinuity'', and focused our attention on papers published since $2015$. We also expanded our search to include other articles that were repeatedly cited by the articles collected in our initial queries.

We also searched in the field of Education, as methodological researchers in that area were among the earliest scholars who discussed the multi-score RD design in non-geographic settings. For this search, we started with the seminal methodological papers by \cite{Reardon-Robinson_2012_JREE} and \cite{Wong-Steiner-Cook_2013_JEBS}, and collected papers that cited these papers or were cited by them. Finally, we also included some papers in environmental science to illustrate uses of the BD design at the intersection between the natural and social sciences. Our review is not meant to provide an exhaustive list of the empirical BD literature. This literature is vast and cannot be fully captured by our approach since we excluded several journals and restricted the time period. Our goal is simply to provide an overview of recent empirical work using the BD design across the social sciences to gauge the most common strategies used for empirical analysis.

The results of our search are summarized in Table \ref{tab:LitReview}. Our search yielded $82$ published papers. Of these, we found that the overwhelming majority ($76$ articles or approximately 93\%) used a pooling approach, focusing on the average effect along the entire boundary and using the univariate closest distance to the boundary as the running variable. Fewer than ten articles were explicit in accounting for the heterogeneity along the boundary; of these, only five reported effects for different points or segments along the boundary \citep{gonzalez2021cell,grout2011land, Keele-Titiunik_2015_PA, snider2015barriers, velez2019tuning}. In sum, our review of the practical literature indicates that most applied researchers adopt a pooling approach, performing localization by grouping all observations sufficiently close to the boundary and then estimating a single average treatment effect. These common practices motivate the taxonomy we use in the rest of this chapter.

\section{Pooling-Based Methods}\label{sec: Pooling-Based Methods}

Given a sample of outcomes and bivariate scores, $(Y_1, \bX_1), \ldots, (Y_n, \bX_n)$, a distance function  $\d(\cdot,\cdot)$ is used to measure the closeness between any two points in $\mathcal{X}$. The most common example is the Euclidean distance $\d(\bx_1, \bx_2) = \|\bx_1 - \bx_2\| = \sqrt{(x_{11} - x_{21})^2 + (x_{12} - x_{22})^2}$ for $\bx_j=(x_{j1},x_{j2})$, $j=1,2$, although in some applications this is not the most appropriate measure. For example, \citet{ambrus2020loss} are interested in how close residences in a city are to each other in terms of how long it takes for a person to walk between them. This requires calculating a walking distance, which is different from the Euclidean distance that cannot take into account, for example, that the path must follow city blocks and cannot go through buildings. More broadly, in geographic settings it is common practice to employ specific geodetic distances \citep{Banerjee_2005_Biometrics}.

Once a distance function $\d(\bX_i, \bx)$ has been chosen, researchers who employ the pooling approach define the closest signed distance to the boundary:
\begin{align*}
    D_i \equiv \big(\Indicator(\bX_i \in \A_1) - \Indicator(\bX_i \in \A_0)\big) \cdot \inf_{\bx \in \B}  \d(\bX_i, \bx),
\end{align*}
for $i=1,\ldots, n$. For a unit $i$, $D_i$ measures how far $i$'s location $\bX_i$ is to the boundary point that is closest to $\bX_i$, regardless of where that boundary point is located. Following the standard RD design logic, but now expanded to the BD design, localization occurs simultaneously along the entire assignment boundary because the univariate distance $D_i$ to $\B$ is used to construct the region such that $|D_i| \leq h$, which covers the boundary $\B$, where $h$ denotes a bandwidth parameter. This region, called a ``tubular neighborhood'' in mathematics \citep{federer2014geometric}, is a natural generalization of the standard localization approach used in univariate RD designs, where the region is simply $|X_i-c|\leq h$ when the score $X_i$ is scalar and $c$ is the univariate cutoff. 

\begin{figure}
    \centering
    \begin{subfigure}[b]{0.45\textwidth}
        \centering
        \includegraphics[width=\linewidth]{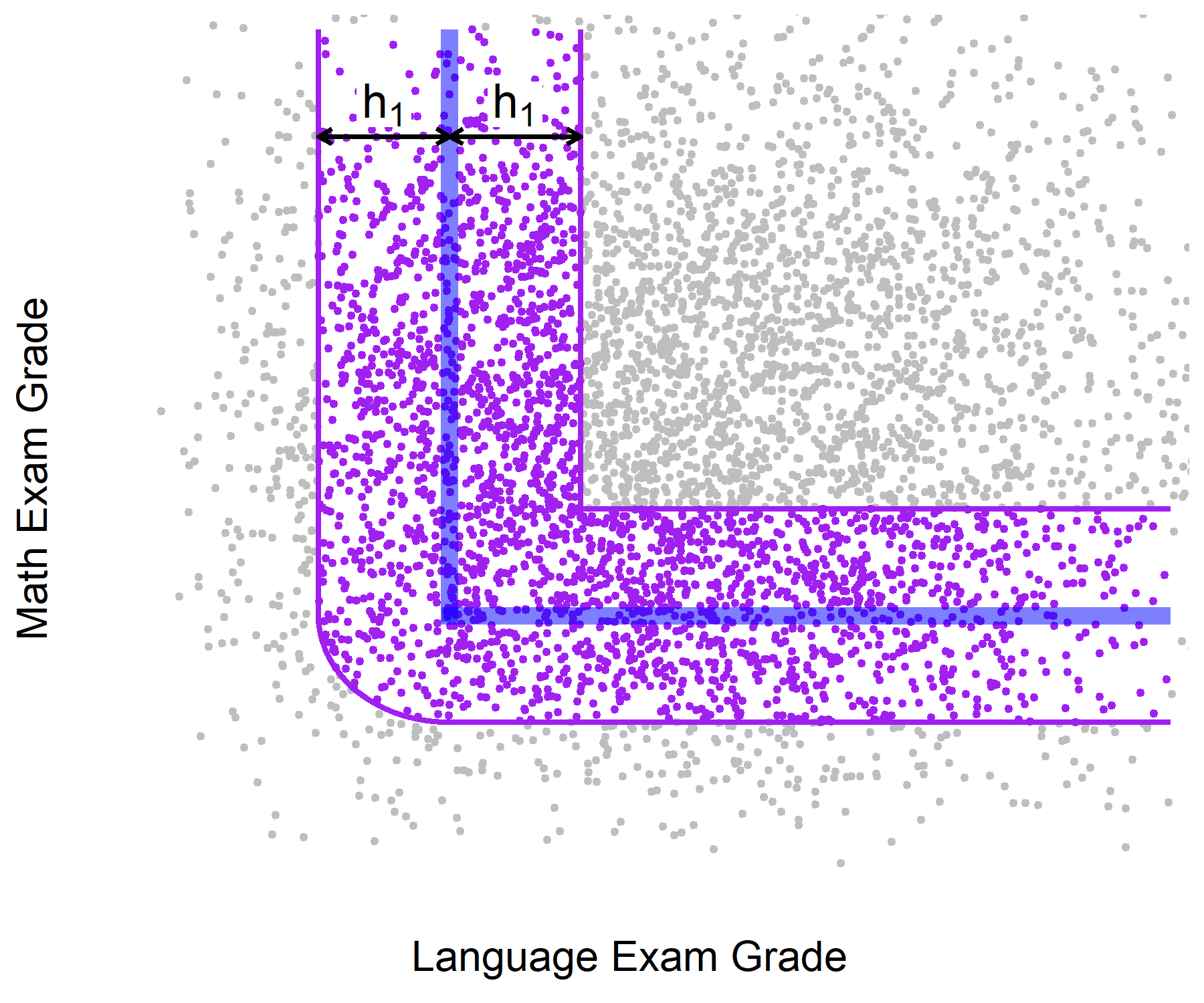}
        \caption{Large Bandwidth}
        \label{fig:Pooling-h1}
    \end{subfigure}
    \hspace{.25in}
    \begin{subfigure}[b]{0.45\textwidth}
        \centering
        \includegraphics[width=\linewidth]{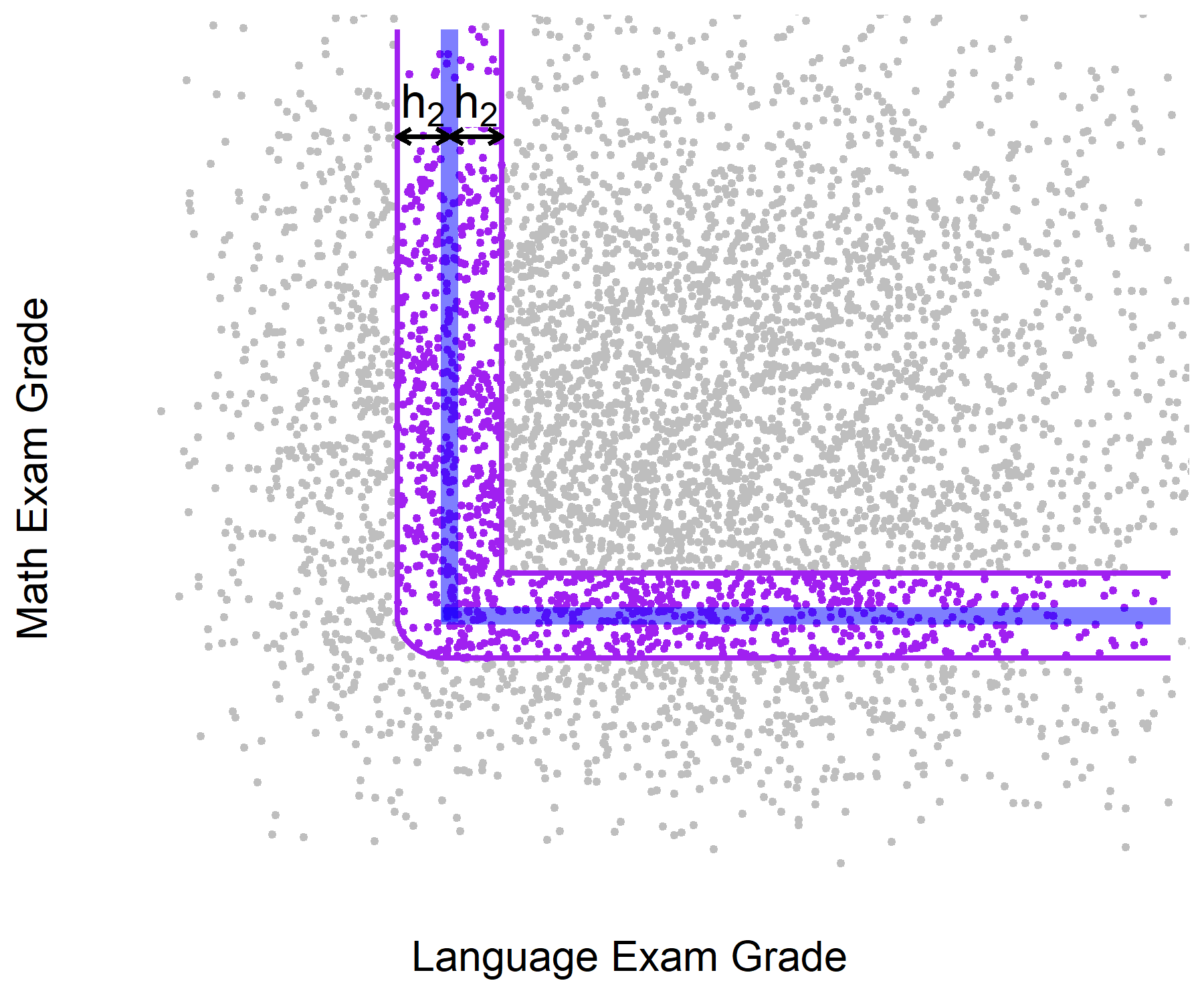}
        \caption{Small Bandwidth}
        \label{fig:Pooling-h2}
    \end{subfigure}    
    \caption{Pooling-Based Method: Localization around Entire Assignment Boundary}
    \label{fig:Pooling}
\end{figure}

Figure \ref{fig:Pooling} illustrates the idea graphically. In both cases, the tubular neighborhood represents the area that includes the observations used in the pooling-based analysis. Figure \ref{fig:Pooling-h1} shows a large bandwidth $h_1$, while Figure \ref{fig:Pooling-h2} shows a smaller bandwidth $h_2$. All units contribute simultaneously, regardless of their specific location. In particular, two units $i$ and $j$ with the same signed distance to the boundary, $D_i=D_j=d$, will be assigned to the same group (treatment or control, depending on the sign of $d$) and will be $|d|$ units away from the boundary, but could be far from each other.

\begin{figure}
    \centering
    \begin{subfigure}[b]{0.45\textwidth}
        \centering
        \includegraphics[width=\linewidth]{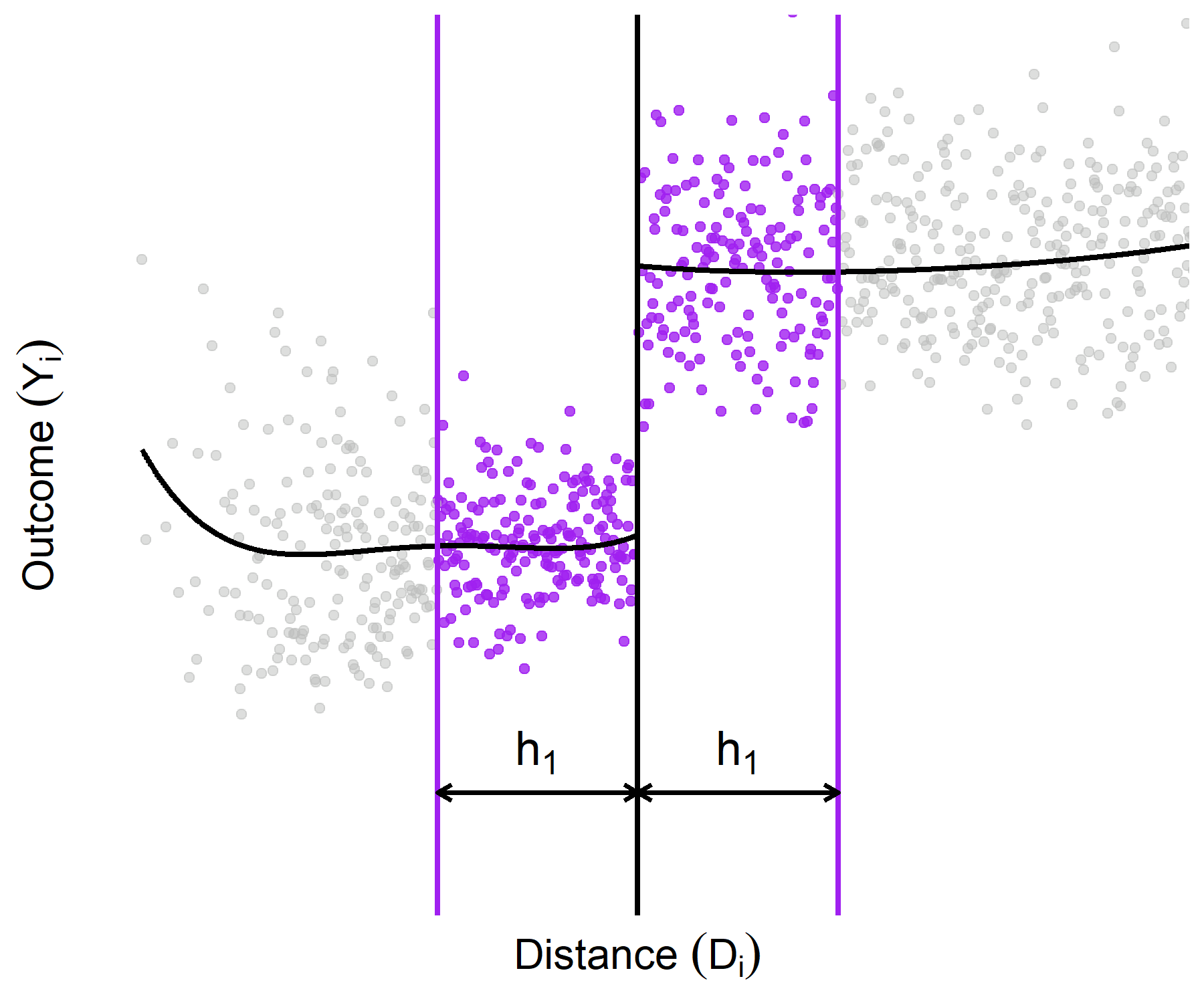}
        \caption{Large Bandwidth}
        \label{fig:Pooling-RDPLOT-h1}
    \end{subfigure}
    \hspace{.25in}
    \begin{subfigure}[b]{0.45\textwidth}
        \centering
        \includegraphics[width=\linewidth]{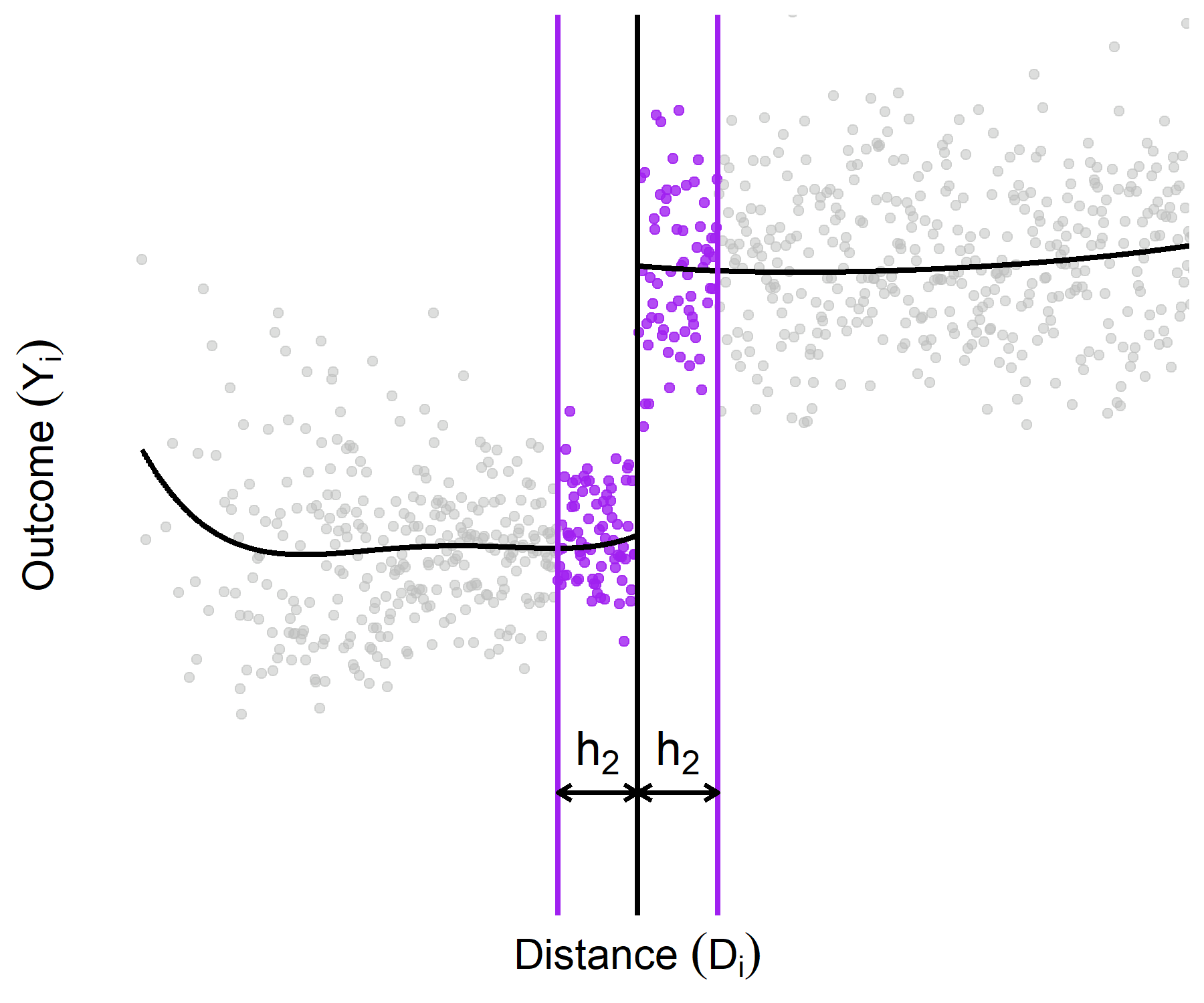}
        \caption{Small Bandwidth}
        \label{fig:Pooling-RDPLOT-h2}
    \end{subfigure}    
    \caption{Pooling-Based Method: RD Plots (with Global Quartic Polynomial Estimates)}
    \label{fig:Pooling-RDPLOT}
\end{figure}

Figure \ref{fig:Pooling-RDPLOT} presents the two resulting RD plots \citep{Calonico-Cattaneo-Titiunik_2015_JASA} in the pooling approach, based on the outcome data $Y_i$ and the univariate distance score $D_i$ (the cutoff is $c=0$ by construction of $D_i$). These plots include a fourth-order global polynomial fit (solid line) for visualization purposes, but these global approximations are not recommended for estimation of treatment effects because they tend to exhibit poor performance due to Runge's phenomenon. Instead, as anticipated in Figure \ref{fig:Pooling}, the idea is to localize around the cutoff (i.e., the one-dimensional boundary curve) to conduct estimation and inference.

In the pooling approach, the analysis is conducted using all the observations with score $\bX_i$ within the tubular neighborhood determined by the bandwidth $h$, that is, for all units with $|D_i|\leq h$. However, within this subsample, the specific treatment effect estimation approach can vary substantially. To describe and unify the most common approaches in the empirical literature illustrated in Table \ref{tab:LitReview} as well as other approaches recommended from a methodological perspective, we define the following generic weighted least squares regression notation:
\begin{align*}
    \reg \quad Y_i \quad \mathtt{on}
         \quad \bZ_i^\top
         \qquad [ \; W_i \; ]
    \qquad\iff\qquad
    \min_{\bbeta} \sum_{i=1}^n \big(Y_i - \bZ_i^\top\bbeta \big)^2 \cdot W_i
\end{align*}
where $Y_i$ is the outcome variable, $\bZ_i$ is a (column) vector of independent variables, and $W_i$ are weights. A localized regression analysis can be implemented by setting $W_i = \Indicator(|D_i|\leq h)$ or, more generally, $W_i = K(D_i/h)$ for a kernel function $K(\cdot)$ determining the relative weight given to each observation. The bandwidth $h$ determines the degree of localization, as illustrated in Figure \ref{fig:Pooling}. When $\bZ_i$ contains polynomial expansions of $D_i$, the resulting estimation approach is often called nonparametric local polynomial regression but, as we will see shortly, empirical researchers often employ other variables in their local fit.

We also define the treatment assignment indicator
\begin{align*}
    T_i = \Indicator(D_i \geq 0) = \Indicator(\bX_i \in \A_1).
\end{align*}
All empirical approaches naturally include $T_i$ in $\bZ_i$ to estimate treatment effects; the majority also include other variables such as $D_i$, $\bX_i$, boundary-segment fixed effects (formally defined below), as well as certain transformations and interactions thereof. For simplicity, we omit discussing the inclusion of predetermined covariates, but we note that researchers often further augment the basic local regression specifications with pretreatment variables such as census characteristics, terrain ruggedness and elevation, time fixed effects, among other features. For a discussion of the role of preintervention covariates in the standard RD design see \cite{Calonico-Cattaneo-Farrell-Titiunik_2019_RESTAT} for efficiency gains  and \cite{Calonico-Cattaneo-Farrell-Palomba-Titiunik_2025_wp} for treatment effect heterogeneity.

The pooling approach for analyzing BD designs begins by choosing the localization bandwidth $h$ to determine the units with $|D_i|\leq h$ that will be used in the subsequent estimation. Until now, there were no formal methods for choosing the bandwidth $h$ in an objective and data-driven way, so researchers often considered a range of ad-hoc values informed by practice and substantive knowledge. For example, \cite{black1999better} reported results in three different bandwidths, keeping observations within $0.35$, $0.25$, and $0.15$ miles of the nearest point on the boundary. As in standard univariate RD designs, different choices of $h$ trade off bias and variance in the estimation, with a larger $h$ leading to more bias but less variance, and a smaller $h$ having the opposite effect. \cite{Cattaneo-Titiunik-Yu_2025_BDD-Pooling} discusses formal results characterizing this tradeoff and provides formal guidance for choosing the bandwidth when implementing pooling BD methods. We discuss these methods in Section \ref{sec: Pooling-Based -- Methodological Results}.

Once the bandwidth $h$ has been chosen, researchers pool all units with distance $D_i$ within $h$ and then compare those assigned to the treatment group ($D_i\geq0$) to those assigned to the control group ($D_i<0$). Because the bandwidth localizes the analysis to only a small region on either side of the boundary $\B$, the empirical analysis compares barely treated units to barely control units which, as in the standard univariate RD design, is the basis for the causal interpretation of the comparison, as all confounders are assumed to vary smoothly at the boundary while the treatment assignment changes abruptly from zero to one.

The most basic analysis proceeds by assuming that the units within the shrinking tubular neighborhood of width $2h$ are as-if randomly assigned to treatment and control. This approach is akin to the ``local randomization'' framework in standard RD designs \citep{Lee-Lemieux_2010_JEL,Cattaneo-Frandsen-Titiunik_2015_JCI,Cattaneo-Titiunik-VazquezBare_2017_JPAM}. A natural treatment effect estimator is the difference-in-means of the outcome between treated and control units with distance to the boundary no larger than the bandwidth: 
\begin{align}\label{eq: reg-Pooling-DIM}
    \reg \quad Y_i \quad \mathtt{on}
         \quad 1, \; T_i
         \qquad [ \; \Indicator( |D_i|\leq h ) \; ],
\end{align}
where the coefficient on $T_i$ is
\begin{align*}
    \widehat{\tau}_{\mathtt{DIM}}
    = \frac{1}{N_1(h)} \sum_{i=1}^n \Indicator(0 \leq D_i \leq h) \cdot Y_i
    - \frac{1}{N_0(h)} \sum_{i=1}^n \Indicator(-h \leq D_i < 0) \cdot Y_i,
\end{align*}
with $N_0(h) = \sum_{i=1}^n \Indicator(-h \leq D_i < 0)$ and $N_1(h) = \sum_{i=1}^n \Indicator(0 \leq D_i \leq h)$. This specification can be interpreted as a local-constant polynomial fit on $D_i$; it does not include the score in the fit because under the assumption of local randomization this score is uncorrelated with the outcome.

\begin{figure}
    \centering
    \begin{subfigure}[b]{0.45\textwidth}
        \centering
        \includegraphics[width=\linewidth]{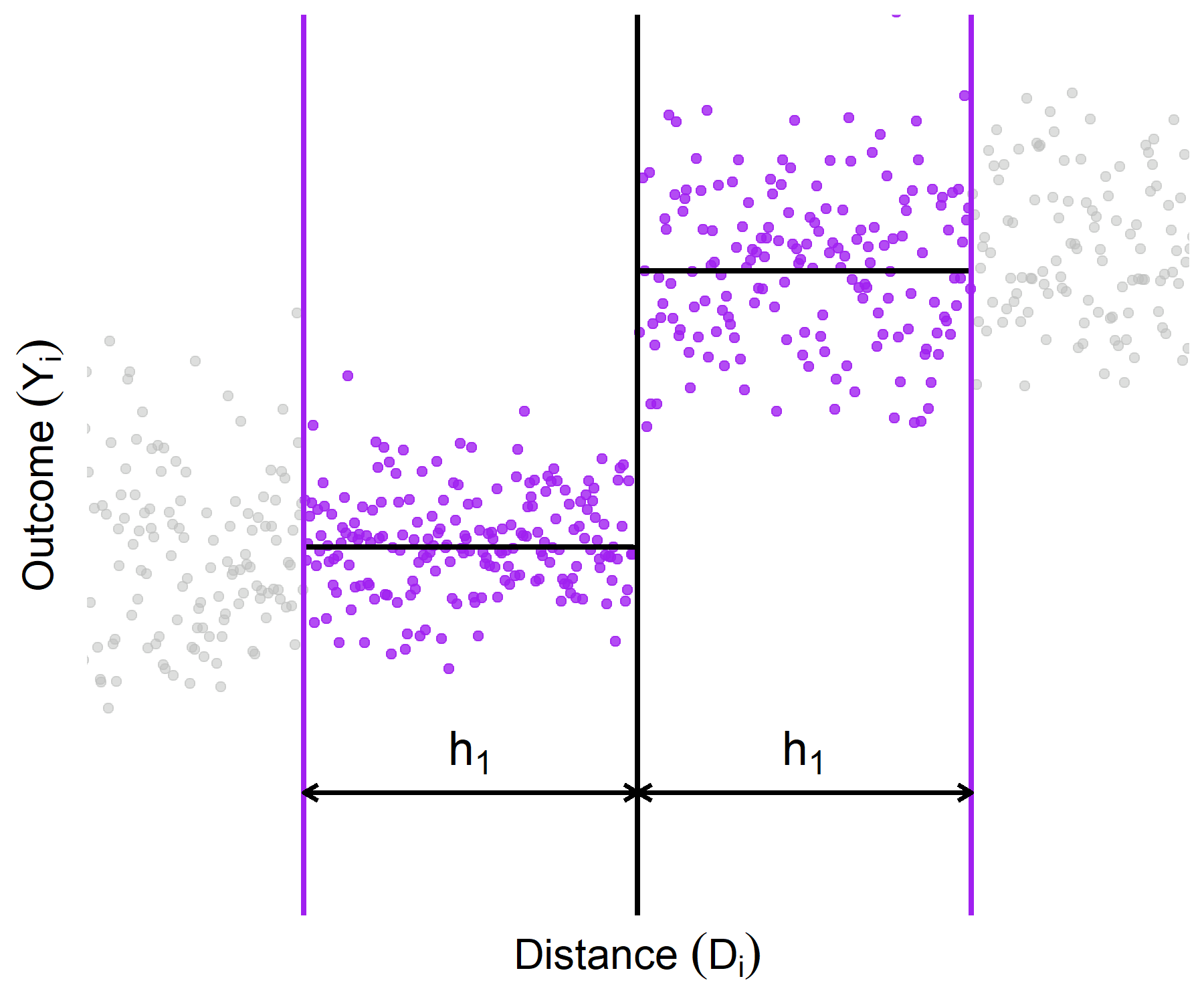}
        \caption{Large Bandwidth}
        \label{fig:Pooling-polyreg-p0-h1}
    \end{subfigure}
    \hspace{.25in}
    \begin{subfigure}[b]{0.45\textwidth}
        \centering
        \includegraphics[width=\linewidth]{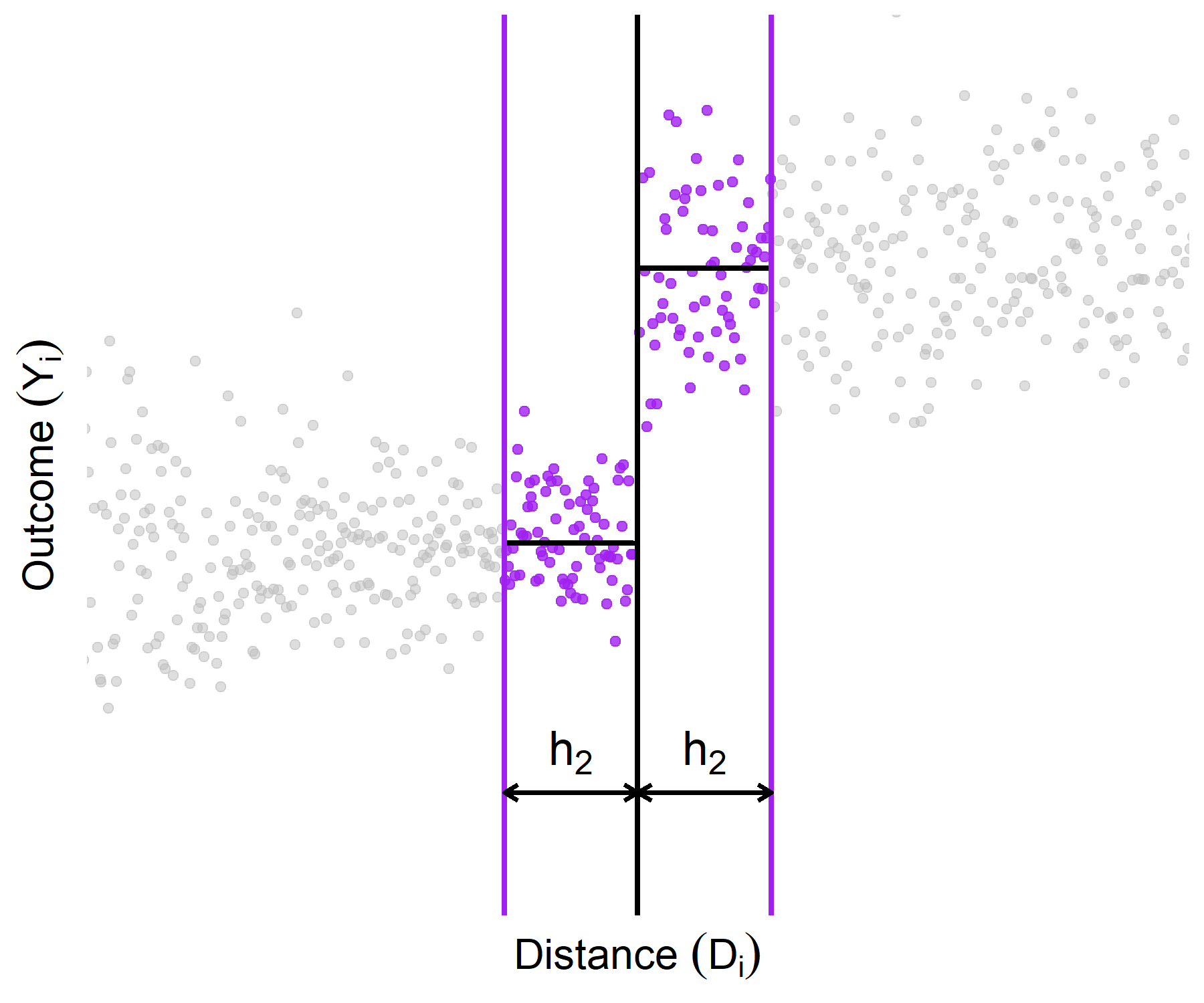}
        \caption{Small Bandwidth}
        \label{fig:Pooling-polyreg-p0-h2}
    \end{subfigure}    
    \caption{Pooling-Based Method: Local-Constant Estimation ($p=0$)}
    \label{fig:Pooling-polyreg-p0}
\end{figure}

\citet[Table III]{black1999better} reported results using specification \eqref{eq: reg-Pooling-DIM}, calling it a ``rough, nonparametric estimate of the value of better schools'' (p. 590). She also augmented this basic specification to include ``segment fixed effects'', that is, binary variables indicating whether an observation is closest to one of several segments partitioning the assignment boundary. To describe this approach formally, we let $\B_{\ell}$, $\ell=1,\ldots,L$, be disjoint segments that partition the assignment boundary, that is, $\B = \sqcup_{1 \leq \ell \leq L} \B_{\ell}$. Then, $S_i = \argmin_{1 \leq \ell \leq L} d(\bX_i, \B_{\ell})$ is the boundary-segment allocation random index for each unit $i=1,\ldots, n$ (when the $\argmin$ contains more than one element, the smallest one is chosen). The vector of segment fixed effects for observation $i$ is $\biota_L(S_i) = (\Indicator(S_i = 1), \cdots, \Indicator(S_i = L))$, which collects the $L$ boundary segment binary indicators determined by $i$'s location. The difference-in-means specification with segment fixed effects is 
\begin{align}\label{eq: reg-Pooling-DIM-SFE}
    \reg \quad Y_i \quad \mathtt{on} 
         \quad \biota_L(S_i), \; T_i
         \qquad [ \; \Indicator( |D_i|\leq h ) \; ],
\end{align}
where the intercept is removed to prevent perfect collinearity. Setting $L=1$ reduces $\biota_L(S_i)$ to the common intercept in specification \eqref{eq: reg-Pooling-DIM}.

\citet[Table II]{black1999better} used a modified version of specification \eqref{eq: reg-Pooling-DIM-SFE} that included the test scores of the school associated with observation $i$ instead of the binary treatment assignment indicator $T_i$. The causal interpretation of this specification rests on the assumption that, after localizing the analysis to observations near the boundary of school attendance districts, the only factor that can explain a difference in average house prices between the treated and control areas is school quality, because school quality changes discontinuously at the boundary while all other determinants of house prices are equal on average. In this sense, using test scores instead of the treatment assignment indicator in specifications \eqref{eq: reg-Pooling-DIM} and \eqref{eq: reg-Pooling-DIM-SFE} is in the spirit of an instrumental variables strategy where $T_i$ is used as an instrument for test scores (which proxy school quality), and the localization to a small bandwidth around the boundary ensures that units assigned to treatment and units assigned to control are similar in all other characteristics that might affect house prices.

The seminal research design used by \cite{black1999better} is prior to the development of modern methods for RD analysis, which occurred predominantly during the $2000$s; see \cite{Cattaneo-Titiunik_2022_ARE} for an overview. In more recent practice, the simple specifications \eqref{eq: reg-Pooling-DIM} and \eqref{eq: reg-Pooling-DIM-SFE} are not commonly used because researchers tend to adopt a continuity-based approach rather than a local randomization approach for interpretation. Under the continuity-based approach, the local constant fit has poor control of the misspecification bias, and hence practitioners prefer more flexible local regression specifications.

Later empirical work using BD designs made more explicit the connection between geographic discontinuities in treatment assignment and the classical RD design, following more closely the methods for univariate continuity-based RD analysis, which emphasized including the RD score in the polynomial specification to allow for arbitrary dependence between score and outcome even within the bandwidth. Although there is considerable diversity in the specifications used in more recent practice, most researchers who include the score adopted one of two main strategies: a polynomial fit of the outcome on the bivariate score $\bX_i$, or a polynomial fit of the outcome on the univariate distance $D_i$. 

In an influential empirical study, \cite{dell2010persistent} augmented specification \eqref{eq: reg-Pooling-DIM-SFE} by including the bivariate score $\bX_i$. She studied the long-run impact of the mita, a forced mining labor system instituted by the Spanish in Peru and Bolivia between $1573$ and $1812$. Dell used the historical boundary that determined which communities were forced to send labor to define treated and control areas, and  studied the effect of the mita on contemporary outcomes measured at the household level, including consumption and childhood stunting. Her main local regression specification was
\begin{align}\label{eq: reg-Pooling-X-SFE}
    \reg \quad Y_i \quad \mathtt{on} 
         \quad \biota_L(S_i), \; T_i, \; \br_p(\bX_i)
         \qquad [ \; \Indicator( |D_i|\leq h ) \; ],
\end{align}
which includes the treatment indicator, boundary segment fixed effects, and a polynomial expansion of the bivariate score $\bX_i$ (with $p=3$ in her case). The estimated coefficient of interest is always the coefficient that accompanies the treatment indicator $T_i$. In our notation, $\br_p(\bu) = (u_1,u_2,u_1^2,u_1u_2,u_2^2,\ldots,u_1^p,u_1^{p-1}u_2,\ldots,u_1 u_2^{p-1},u_2^p)$ with $\bu=(u_1,u_2)$. The inclusion of (a polynomial basis of) the bivariate location score $\bX_i$ is meant to control for ``the smooth effects of geographic location'' \citet[p. 1875]{dell2010persistent}. A notable characteristic of this specification is that it does not include interactions between the bivariate score and the treatment indicator, which imposes the restriction that the polynomials be identical in the treated and control areas. For bandwidth, Dell used $100$ km, $75$ km and $50$ km within the mita boundary.

\begin{figure}
    \centering
    \begin{subfigure}[b]{0.45\textwidth}
        \centering
        \includegraphics[width=\linewidth]{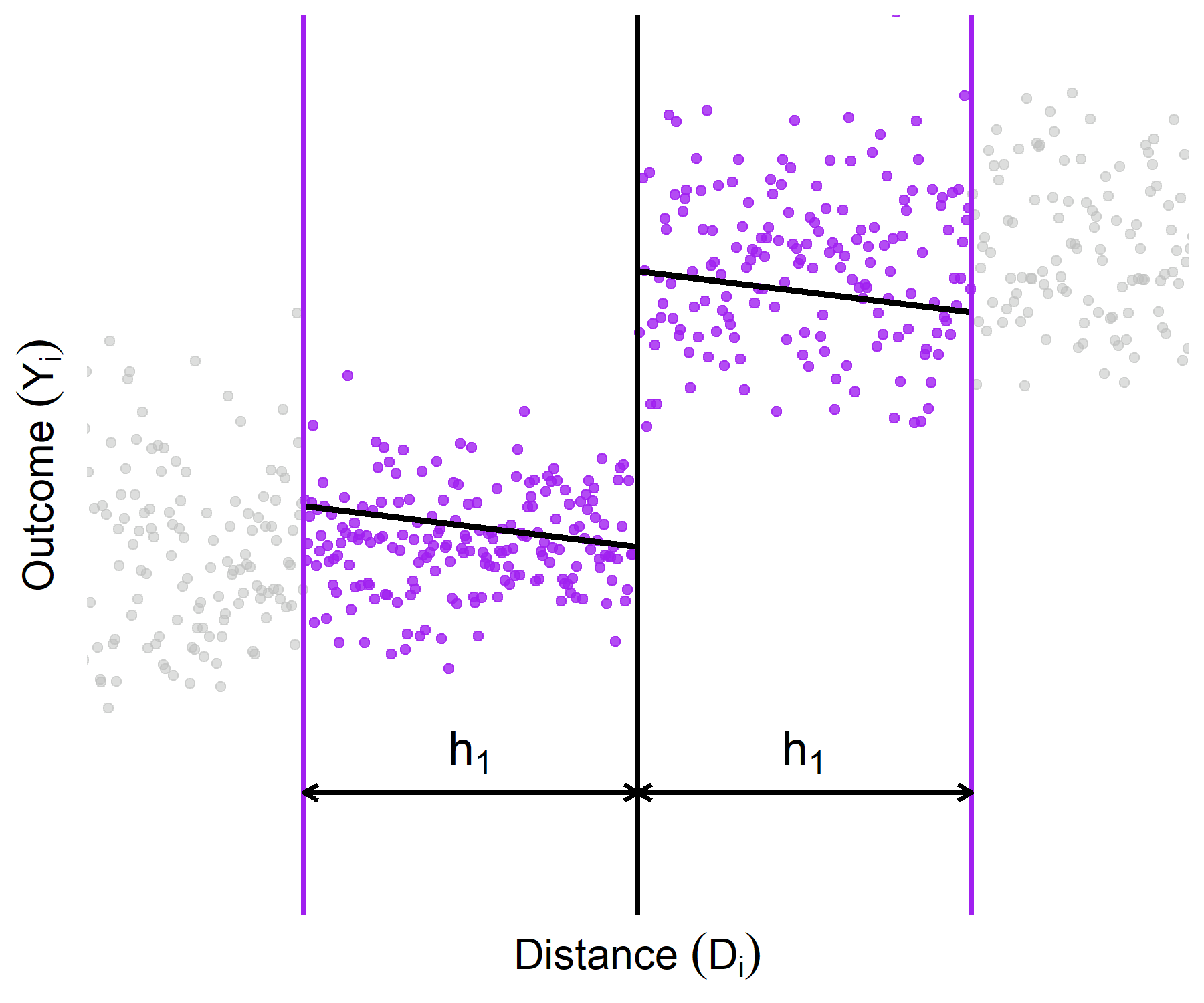}
        \caption{Large Bandwidth}
        \label{fig:Pooling-polyreg-noint-p1-h1}
    \end{subfigure}
    \hspace{.25in}
    \begin{subfigure}[b]{0.45\textwidth}
        \centering
        \includegraphics[width=\linewidth]{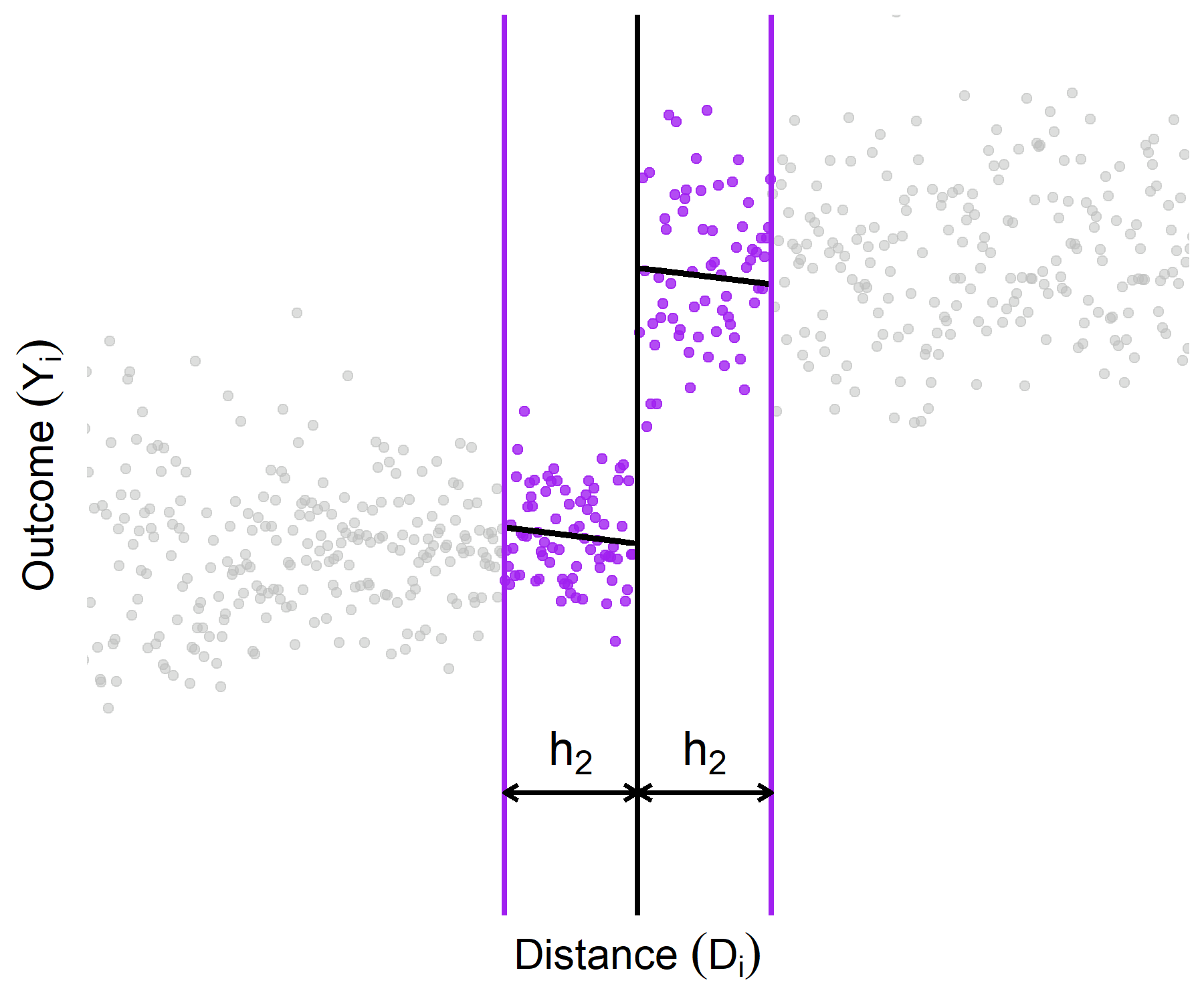}
        \caption{Small Bandwidth}
        \label{fig:Pooling-polyreg-noint-p1-h2}
    \end{subfigure}    
    \caption{Pooling-Based Method: Local-Linear No-Interaction Estimation ($p=1$)}
    \label{fig:Pooling-polyreg-noint-p1}
\end{figure}

Specification \eqref{eq: reg-Pooling-X-SFE} has become a common strategy to analyze BD designs. \cite{mendez2022multinationals} use it to study the effect of private sector companies in the development of local amenities, using a land concession to the multinational United Fruit Company in Costa Rica. \cite{paulsen2023foundations} use it to estimate the effect of school funding on future earnings and political participation in New York, comparing areas that received higher funding to neighboring areas that did not. \cite{de2018agents} compare areas in South Africa where traditional leaders remain highly influential to areas where they are not, and estimate the electoral impact of strong political ties between traditional leaders and the African National Congress (ANC). Other examples are listed in Table \ref{tab:LitReview}. All of these examples follow specification \eqref{eq: reg-Pooling-DIM-SFE} by including a flexible polynomial expansion of latitude and longitude, as well as boundary-segment fixed effects, but without allowing for interactions with the treatment indicator $T_i$.

An alternative approach is to directly mimic the standard univariate RD design, taking the scalar (signed) distance to the closest point on the boundary for every unit, $D_i$, as the univariate score, and employing local polynomial regression. The most basic local regression specification in this context takes the form:
\begin{align}\label{eq: reg-Pooling-D}
    \reg \quad Y_i \quad \mathtt{on}
         \quad 1, \; T_i, \; \br_p(D_i)
         \qquad [ \; \Indicator( |D_i|\leq h ) \; ],
\end{align}
with the polynomial expansion reducing to the simpler vector of the univariate distance $\br_p(D_i) = (D_i,D_i^2,\ldots,D_i^p)$. As in specification \eqref{eq: reg-Pooling-X-SFE}, specification \eqref{eq: reg-Pooling-D} can also be augmented with boundary-segment fixed effects:
\begin{align}\label{eq: reg-Pooling-D-SFE}
    \reg \quad Y_i \quad \mathtt{on}
         \quad \biota_L(S_i), \; T_i, \; \br_p(D_i)
         \qquad [ \; \Indicator( |D_i|\leq h ) \; ].
\end{align}
In both cases, the treatment effect estimator is the coefficient estimate accompanying $T_i$. Figure \ref{fig:Pooling-polyreg-noint-p1} offers a graphical representation of this empirical strategy.

Although specifications \eqref{eq: reg-Pooling-D} and \eqref{eq: reg-Pooling-D-SFE} aim to mimic the univariate RD design, they differ from best practices in an important respect. In the RD design, the recommended specification is a polynomial fit of the outcome on an intercept, the treatment assignment indicator, a polynomial expansion of the score, and the interaction between this expansion and the treatment assignment indicator. The latter interaction is crucial to allow the polynomial fit for control units to be different from the polynomial fit for treated units. None of the specifications discussed so far allow for this flexibility, potentially introducing a larger misspecification bias. 

\begin{figure}
    \centering
    \begin{subfigure}[b]{0.45\textwidth}
        \centering
        \includegraphics[width=\linewidth]{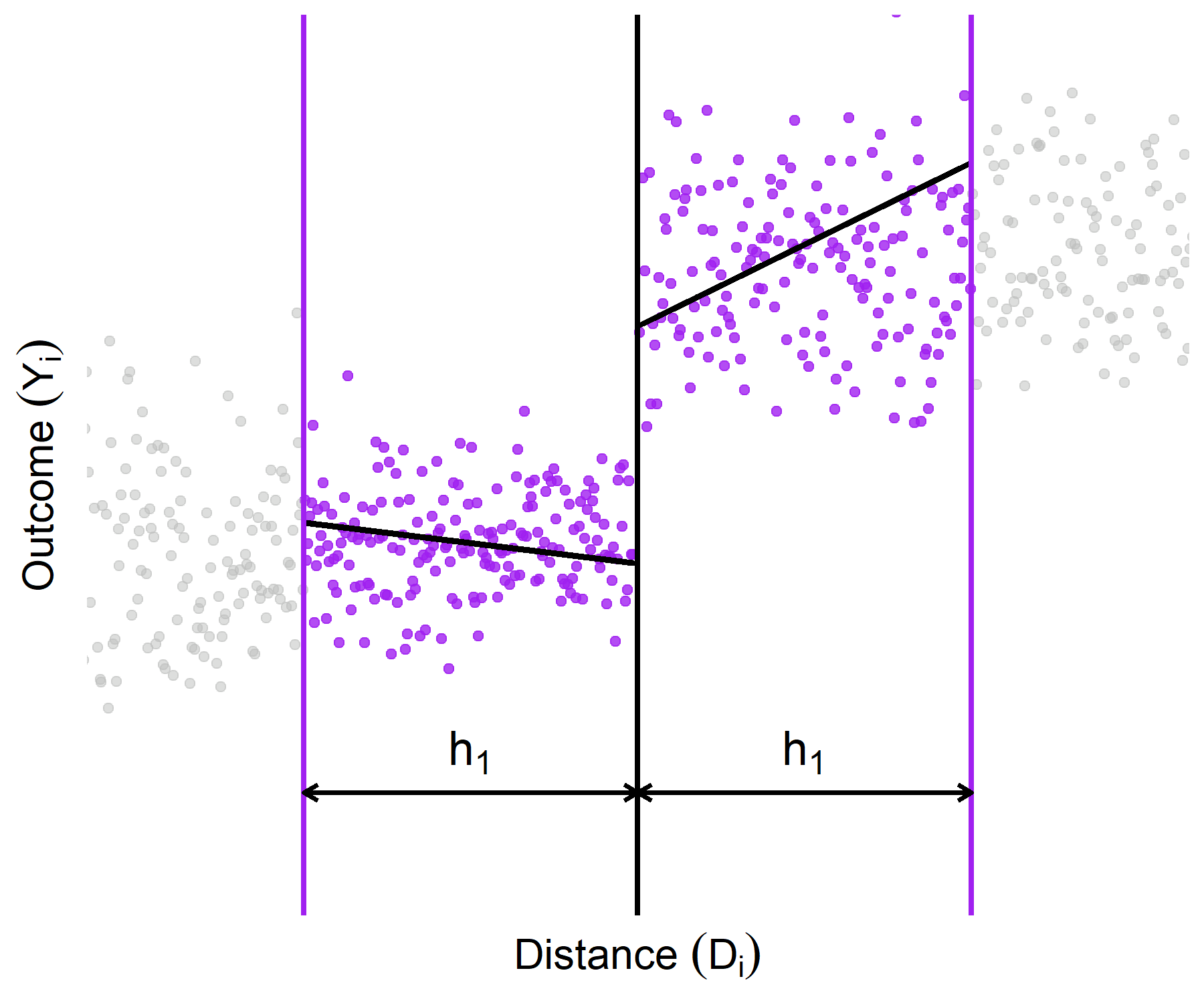}
        \caption{Large Bandwidth}
        \label{fig:Pooling-polyreg-p1-h1}
    \end{subfigure}
    \hspace{.25in}
    \begin{subfigure}[b]{0.45\textwidth}
        \centering
        \includegraphics[width=\linewidth]{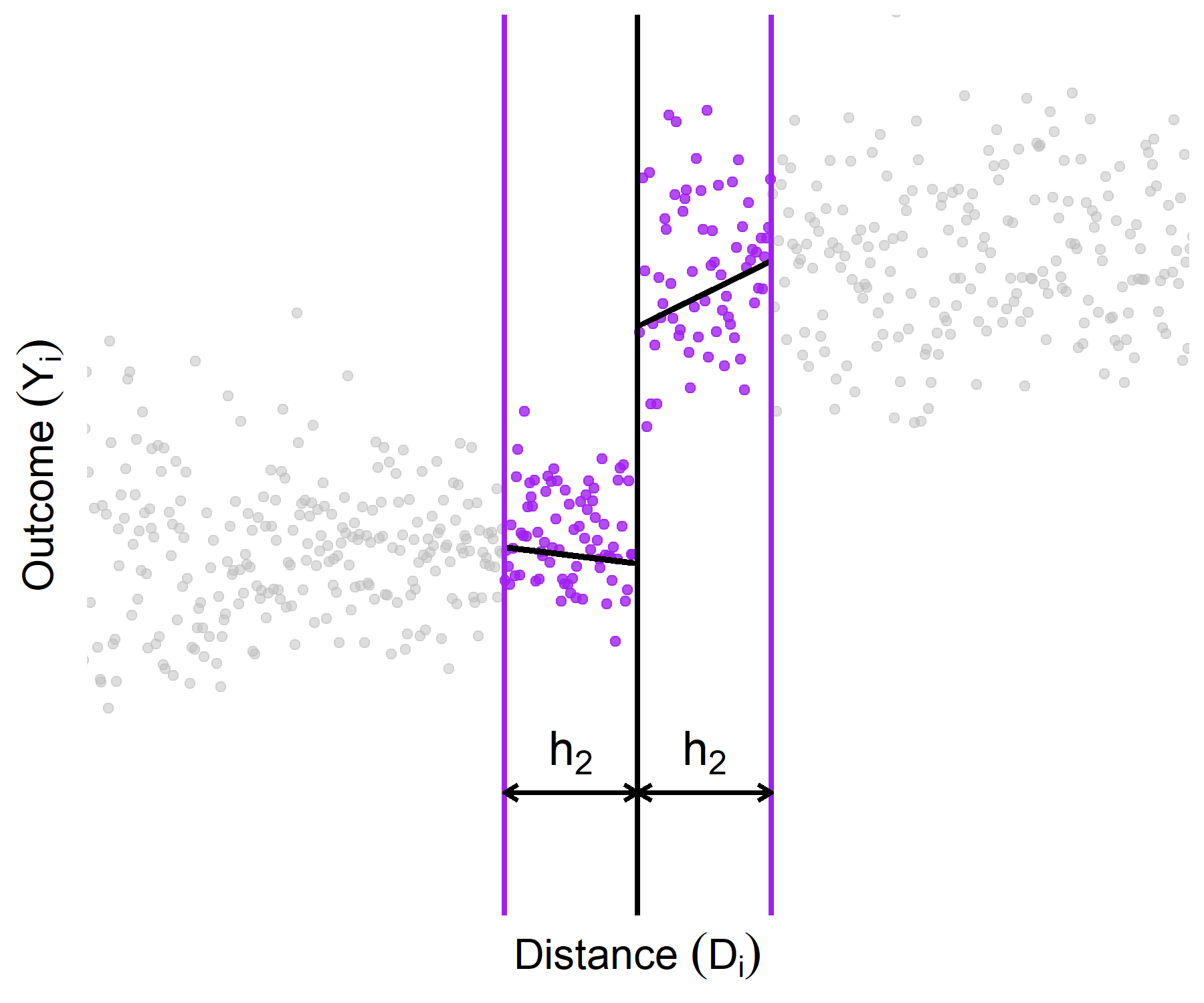}
        \caption{Small Bandwidth}
        \label{fig:Pooling-polyreg-p1-h2}
    \end{subfigure}    
    \caption{Pooling-Based Method: Local-Linear Estimation ($p=1$)}
    \label{fig:Pooling-polyreg-p1}
\end{figure}

The most flexible specification based on $D_i$ takes the form:
\begin{align}\label{eq: reg-Pooling-D-SFE-INTER}
    \reg \quad Y_i \quad \mathtt{on}
         \quad \biota_L(S_i), \; T_i, \; \br_p(D_i), \; T_i\cdot\br_p(D_i)
         \qquad [ \; \Indicator( |D_i|\leq h ) \; ],
\end{align}
where the estimated coefficient of interest continues to be the one accompanying $T_i$. In practice, specification \eqref{eq: reg-Pooling-D-SFE-INTER} is often implemented with $p=1$ and, depending on whether standard software for RD design analysis is used, may employ a non-uniform weighting kernel (e.g., a triangular kernel that down-weights observations as their distance from the boundary increases). Putting aside the inclusion of boundary-segment fixed effects, Figure \ref{fig:Pooling-polyreg-p1} illustrates the importance of including the interaction term $T_i\cdot\br_p(D_i)$. Comparing Figures \ref{fig:Pooling-polyreg-noint-p1} and \ref{fig:Pooling-polyreg-p1} shows that the two approaches can lead to different local approximations, with specifications \eqref{eq: reg-Pooling-DIM} through \eqref{eq: reg-Pooling-D-SFE} potentially exhibiting a larger bias than specification \eqref{eq: reg-Pooling-D-SFE-INTER}.

Specification \eqref{eq: reg-Pooling-D-SFE-INTER} is one of the most commonly used in recent practice. Examples include \citet{ayres2021environmental}, \cite{eugster2017culture}, \citet{jones2022factor}, \citet{lalive2008extended}, and \citet{lowes2021concessions}. We note that several articles describe their empirical specifications as \eqref{eq: reg-Pooling-D} or \eqref{eq: reg-Pooling-D-SFE}, but implement specification \eqref{eq: reg-Pooling-D-SFE-INTER} in their analyses (as clarified in text, footnotes or replication files). These researchers employ standard practices and software from the univariate RD design literature, which follows specification \eqref{eq: reg-Pooling-D-SFE-INTER}.

Specification \eqref{eq: reg-Pooling-X-SFE} can also be augmented to include the interaction term between the treatment assignment and the polynomial expansion of the bivariate score $\bX_i$:  
\begin{align}\label{eq: reg-Pooling-X-SFE-INTER}
    \reg \quad Y_i \quad \mathtt{on}
         \quad \biota_L(S_i), \; T_i, \; \br_p(\bX_i), \; T_i \cdot \br_p(\bX_i)
         \qquad [ \; \Indicator( |D_i|\leq h ) \; ],
\end{align}
where the estimated treatment effect coefficient accompanying $T_i$ enjoys better bias properties due to the added flexibility in the local regression specification. However, this more flexible approach is not as common in the empirical literature---some exceptions are \citet{gonzalez2021cell}, \citet{egger2015impact}, and \citet{jones2022effects}.

In other applications, researchers estimate treatment effects in BD designs by considering two separate univariate RD designs, that is, keeping one dimension fixed and studying the other dimension of the bivariate score. For example, \cite{LondonoVelezRodriguezSanchez_2020_AEJ} investigate the effects of \textit{Ser Pilo Paga} (SPP), a government subsidy in Colombia that provided tuition support for post-secondary students to attend a government-certified higher education institution. Eligibility was based on both merit and economic need: in order to qualify for the program, students had to obtain a high grade in Colombia's national standardized high school exit exam, \textit{SABER 11}, and they also had to come from economically disadvantaged families, as measured by the survey-based wealth index \textit{SISBEN}. The resulting BD design is as illustrated in Figure \ref{fig:BDnongeo} with $\bX_i= (X_{1i},X_{2i}) = (\text{\textit{SABER11}}_i,\text{\textit{SISBEN}}_i)$. The authors analyzed it by considering each cutoff separately and then pooling all observations in the other dimension, leading to two univariate RD designs: one with score $\textit{SISBEN}_i$ in the sample $\text{\textit{SABER11}}_i\geq0$, and the other with score $\text{\textit{SABER11}}_i$ in the sample $\textit{SISBEN}_i\geq0$. This approach is equivalent to splitting the boundary along each of its two linear segments and then employing a pooling  approach within each subsample. This idea leads to the following local regression specification:
\begin{align}\label{eq: reg-Pooling-D-SFE-INTER-SEGHET}
    \reg \quad Y_i \quad \mathtt{on} 
         \quad T_i\cdot\biota_L(S_i), \; \biota_L(S_i)\otimes\br_p(D_i), \; T_i\cdot \biota_L(S_i)\otimes\br_p(D_i)
         \qquad [ \; \Indicator( |D_i|\leq h ) \; ],
\end{align}
where the indicator variable $T_i$ is removed due to the perfect collinearity with $T_i\cdot\biota_L(S_i)$, since $\biota_L(S_i)$ contains indicator variables capturing a partitioning of $\B$, and $\otimes$ denotes the Kronecker product. The treatment effect of interest is now the collection of coefficients accompanying $T_i\cdot\biota_L(S_i)$, one for each segment. Specification \eqref{eq: reg-Pooling-D-SFE-INTER-SEGHET} is roughly equivalent to fitting specification \eqref{eq: reg-Pooling-D-SFE-INTER} $L$ times, in each case only using the data corresponding to one of $L$ segments---the only  difference is whether observations are assigned to a single segment or reused). \cite{ou2010leave}, \cite{LondonoVelezRodriguezSanchez_2020_AEJ} and \cite{salti2022impact} employ this boundary-segment specific pooling-based approach with $L=2$, analyzing each of the two linear segments as in Figure \ref{fig:BDnongeo}.

Some researchers have also included both the univariate distance $D_i$ and the bivariate score $\bX_i$ in their local regression specifications---see, for example, \cite{ehrlich2018persistent} and \cite{dehdari2022origins}. It is also possible to conceptualize more flexible specifications by including other interaction terms between $T_i$, $\br_p(\bX_i)$, and $\biota_L(S_i)$ analogous to specification \eqref{eq: reg-Pooling-D-SFE-INTER-SEGHET}. Our literature review in Table \ref{tab:LitReview} suggests that these approaches are uncommon in practice, so we do not discuss them further.

Although we have not systematically categorized the bivariate score in each of the empirical applications in Table \ref{tab:LitReview}, we suspect that some of the specification choices in practice may be determined by data availability and the granularity of the score. For example, \cite{dell2010persistent} uses latitude and longitude of districts as the bivariate score, but her unit of observation is individuals, creating equal score values or mass points for all individuals in the same district (mass points). When limited data availability prevents the collection of continuously distributed bivariate scores, the statistical properties and methods we discuss in the upcoming sections do not apply directly, as they all assume that the score is a continuously distributed random variable. \cite{Keele-Titiunik_2015_PA, Keele-Titiunik_2016_PSRM} discuss the importance of score granularity in geographic BD designs.

\subsection{Methodological Results}\label{sec: Pooling-Based -- Methodological Results}

While there has been a proliferation of different empirical strategies relying on the pooling of observations closest to the treatment assignment boundary, little is known about their properties and relative merits for the analysis of BD designs. \cite{Cattaneo-Titiunik-Yu_2025_BDD-Pooling} recently investigated pooling-based methods, offering formal identification, estimation, and inference results. Their methodological results rely on techniques from geometric measure theory \citep{federer2014geometric}, which are used to characterize how the different treatment effect estimates emerging from specifications \eqref{eq: reg-Pooling-DIM} through \eqref{eq: reg-Pooling-D-SFE-INTER} behave in large samples, that is, when $h\to0$ as $n\to \infty$. These asymptotic approximations form the basis of the continuity-based approach in standard univariate RD designs \citep{Hahn-Todd-vanderKlaauw_2001_ECMA}, and are nowadays the most common way of characterizing the properties of causal treatment effect estimators in all RD settings. This section summarizes some of the findings in \cite{Cattaneo-Titiunik-Yu_2025_BDD-Pooling}.

The main challenge of pooling-based BD methods is that the localization is done using a sequence of shrinking tubular neighborhoods covering the one-dimensional submanifold $\B$, as illustrated in Figure \ref{fig:Pooling}. An important implication is that the geometry of the treatment assignment boundary and the specific distance function used can substantially affect statistical properties. Our upcoming discussion omits most technicalities and focuses instead on conceptual ideas to aid empirical researchers interpret and implement pooling-based methods.

Because the treatment assignment boundary $\B$ could be highly irregular, particularly in Geographic RD designs, it is necessary to restrict its geometry to establish formal properties. A minimal assumption commonly used in mathematics is to require $\B$ to be a rectifiable curve, that is, to assume that it has finite length (more formally, $\B$ is a rectifiable curve if and only if it can be reparameterized by a Lipschitz continuous function). This restriction allows for formally defining and computing integrals of functions along the one-dimensional domain $\B$. More importantly, under additional technical conditions, regularity on the geometry of $\B$ allows for those integrals to be computed over shrinking tubular neighborhoods and for limits to be well-defined, which is at the core of the results in \cite{Cattaneo-Titiunik-Yu_2025_BDD-Pooling}. Under regularity conditions, the authors establish the following result:
\begin{align}\label{eq: Integral convergence}
    \lim_{h\downarrow0}\int_{\mathcal{T}(h)} \frac{1}{h} g \Big(\frac{\d(\bx, \B)}{h}\Big) m(\bx) \text{d}\bx
    = \mathtt{c}_{\B} \cdot \int_0^1 g(s) ds \cdot \int_{\B} \frac{m(\bx)}{\Jacob \d(\bx, \B)} \text{d}\Haus(\bx),
\end{align}
where $\mathcal{T}(h) = \big\{ \bx\in\X : \d(\bx,\B) \leq h \big\}$, for each $h\geq0$, is a tubular neighborhood covering $\B$,
\begin{align*}
    \d(\bx, \B) = \inf_{\bb \in \B}  \d(\bx, \bb)
\end{align*}
and $\Jacob \d(\bx, \B)$ denotes its Jacobian, and $\Haus(\bx)$ denotes the $1$-dimensional Hausdorff measure. The constant $\mathtt{c}_{\B}$ can also be characterized, but it is not important for the upcoming discussion (it can be set to $\mathtt{c}_{\B}=2$ for the analysis of pooling-based methods). Note that $D_i = \big(\Indicator(\bX_i \in \A_1) - \Indicator(\bX_i \in \A_0)\big) \cdot \d(\bX_i,\B)$.

The integral $\int_{\B} \frac{m(\bx)}{\Jacob \d(\bx, \B)} \text{d}\Haus(\bx)$ is a natural generalization of the standard line integral, where $\text{d}\Haus$ serves as the rigorous ``length element'', enabling a robust framework for integration over complex one-dimensional domains in the plane. In particular, if $\B$ is piecewise linear (as in Figures \ref{fig:BDnongeo} and \ref{fig:Pooling}), the integral can be reduced to the sum of integrals over the linear segments using a natural smooth curve parametrization and standard Riemann integration (when $m(\bx)$ is Riemann integrable). For notational simplicity, we write
\begin{align*}
    \int_{\B} m(\bb) \text{d}\bx = \int_{\B} \frac{m(\bx)}{\Jacob \d(\bx, \B)} \text{d}\Haus(\bx),
\end{align*}
but with the understanding that the left-hand-side integral is just notation for the rigorously defined right-hand-side integral.

The limiting integral in \eqref{eq: Integral convergence} can be used to give precise meaning to the probability limit of the treatment effect estimates from any of the pooling specifications discussed previously. However, we need to introduce standard potential outcomes notation \citep[see, e.g.,][]{Hernan-Robins_2020_Book} to give a causal interpretation to the probability limit of those parameters. For $i = 1,2,\dots, n$, let $Y_i(0)$ and $Y_i(1)$ denote the potential outcomes for unit $i$ under control and treatment assignment, respectively. The  observed outcome is
\begin{align*}
    Y_i = (1 - T_i) \cdot Y_i(0) + T_i \cdot Y_i(1).
\end{align*}
To streamline the presentation, we  illustrate the methodological findings in \cite{Cattaneo-Titiunik-Yu_2025_BDD-Pooling} using the simplest specification \eqref{eq: reg-Pooling-DIM}, leading to the local difference-in-means estimator $\widehat{\tau}_{\mathtt{DIM}}$. The other more flexible local regression specifications enjoy similar properties.

The probability limit of $\widehat{\tau}_{\mathtt{DIM}}$ can be naturally characterized as a ratio of integrals along the boundary. As $n\to\infty$ and leveraging the logic of the law of large numbers, $\frac{1}{nh} N_1(h) = \frac{1}{nh} \sum_{i=1}^n \Indicator(0 \leq D_i \leq h)$ will be close in probability to $\frac{1}{h}\E[\Indicator(0 \leq D_i \leq h)]$, and
\begin{align*}
    \frac{1}{h}\E[\Indicator(0 \leq D_i \leq h)]
    = \int_{\mathcal{T}_1(h)} \frac{1}{h} g \Big(\frac{\d(\bx, \B)}{h}\Big) f(\bx) \text{d}\bx
    \to \int_0^1 g(s) ds \cdot \int_{\B} f(\bx) \text{d}\bx,
\end{align*}
where the limit is taken as $h\to0$, $f(\bx)$ denotes the Lebesgue density of $\bX_i$, $\mathcal{T}_1(h) = \big\{ \bx\in\X : 0 \leq D_i \leq h \big\}$, and the conclusion follows by \eqref{eq: Integral convergence} upon setting $g(u) = \Indicator(0 \leq u < 1)$ and $m(\bx) = \Indicator(\bx\in\A_1) f(\bx)$. While $g(u)$ is redundant in this case, and $\int_0^1 g(s) ds=1$, we nonetheless make it explicit to highlight that a term will appear when a non-uniform kernel is used. Similarly, $N_0(h)/(nh) \to_\P \int_{\B} f(\bx) \text{d}\bx$. Furthermore, by the same logic, the two (properly rescaled) numerators of $\widehat{\tau}_{\mathtt{DIM}}$ will be close in probability to their expectations, where
\begin{align*}
    \frac{1}{h}\E[\Indicator(0 \leq D_i \leq h) \cdot Y_i]
    &= \frac{1}{h}\E\Big[\Indicator(0 \leq D_i \leq h) \cdot \E[ Y_i(1) | D_i ] \Big]\\ 
    &= \int_{\mathcal{T}_1(h)} \frac{1}{h} g \Big(\frac{\d(\bx, \B)}{h}\Big) \E[ Y_i(1) | \d(\bX_i, \B) = \d(\bx, \B) ] f(\bx)  \text{d}\bx\\
    &\to \int_0^1 g(s) ds \cdot \int_{\B} \E[Y_i(1) | \bX_i=\bx] f(\bx) \text{d}\bx,
\end{align*}
and, analogously,
\begin{align*}
    \frac{1}{h}\E[\Indicator(-h \leq D_i \leq 0) Y_i]
    \to \int_0^1 g(s) ds \cdot \int_{\B} \E[Y_i(0) | \bX_i=\bx] f(\bx) \text{d}\bx.
\end{align*}
The above calculations require precise regularity conditions and technical work, including restrictions on the geometry of $\B$ and the distance function $\d(\cdot)$ to guarantee that the expressions are well-defined and the limits are valid.

Putting the above calculations together,
\begin{align*}
    \widehat{\tau}_{\mathtt{DIM}}
    \to_\P \tau = \frac{\int_\B \tau(\bx) f(\bx) \text{d}\bx}{\int_\B f(\bx) \text{d}\bx},
    \qquad
    \tau(\bx) = \E[ Y_i(1) - Y_i(0) | \bX_i=\bx].
\end{align*}
The parameter $\tau$ is called the Boundary Average Treatment Effect (BATE) by \cite{Cattaneo-Titiunik-Yu_2025_BDD-Pooling}. This parameter was heuristically introduced in the education literature by \cite{Wong-Steiner-Cook_2013_JEBS}, who expressed it as
\begin{align*}
    \tau = \int_\B \tau(\bx) f(\bx|\bX_i\in\B) \text{d}\bx = \E[Y_i(1) - Y_i(0) | \bX_i=\bx , \bX_i\in\B],
\end{align*}
using the notation $f(\bx|\bX_i\in\B) = \frac{f(\bx)}{\int_\B f(\bx) \text{d}\bx}$. Around the same time, \cite{Keele-Titiunik_2015_PA} also discussed the BATE parameter in the context of Geographic RD designs.

The parameter $\tau$ captures a density-weighted average of average treatment effects at each point along the boundary, and thus aggregates the potentially heterogeneous treatment effects captured by $(\tau(\bx):\bx\in\B)$. \cite{Cattaneo-Titiunik-Yu_2025_BDD-Location} call $\tau(\bx)$ the Boundary Average Treatment Effect Curve (BATEC). As we discuss in Section \ref{sec: Distance-based and Location-Based Methods}, this functional causal parameter can be seen as a building block for several other causal parameters of interest in BD designs.

Going beyond identification, \cite{Cattaneo-Titiunik-Yu_2025_BDD-Pooling} study estimation accuracy and uncertainty quantification for pooling-based methods. For point estimation, the authors shows that
\begin{align*}
    |\widehat{\tau}_{\mathtt{DIM}} - \tau|^2 \approx_\P \mathsf{B}_n^2 + \frac{\mathsf{V}}{nh},
\end{align*}
when $h\to0$ as $n\to\infty$, and where $\approx_\P$ denotes approximation in probability up to higher order terms, $\mathsf{B}_n$ denotes a leading bias term, and $\mathsf{V}$ denotes the asymptotic variance that is bounded and bounded away from zero. Characterizing the precise order of the bias is difficult without additional restrictions on the treatment assignment boundary $\B$: the authors show that $\mathsf{B}_n = O(h)$ in general. Furthermore, as in the case of standard univariate RD designs, and under additional regularity conditions, the order of the bias may be improved when $\br_p(D_i)$ and $T_i \cdot \br_p(D_i)$ are included in the local regression specification, as in \eqref{eq: reg-Pooling-D-SFE-INTER}, leading to $\mathsf{B}_n = O(h^{p+1})$. Another interesting feature is that the rate of decay of the variance component is the same as in one-dimensional nonparametric estimation: this improvement in convergence rate comes from the aggregation along the boundary of the bivariate nonparametric function $\tau(\bx)$. See \cite{Cattaneo-Titiunik-Yu_2025_BDD-Location} and \cite{chen2025semiparametric} for more discussion.

For bandwidth selection, under regularity conditions on $\B$ and $\d(\cdot)$, the bias $\mathsf{B}_n$ may admit a valid expansion as in the standard RD design literature. Therefore, optimal bandwidth selection procedures based on mean square error (MSE) or coverage error minimization may be used for implementation of pooling methods. See \cite{Calonico-Cattaneo-Farrell_2020_ECTJ} for a review.

For distribution theory and inference, consider the usual test statistic
\begin{align*}
    \widehat{\T} = \frac{\widehat{\tau}_{\mathtt{DIM}} - \tau}{\sqrt{\widehat{\mathbb{V}}[\widehat{\tau}_{\mathtt{DIM}}]}},
\end{align*}
where $\widehat{\mathbb{V}}[\widehat{\tau}_{\mathtt{DIM}}]$ denotes any of the usual variance estimators from (local, weighted) least squares regression methods. Under regularity conditions, and if $n h \to \infty$ and $nh \mathsf{B}_n^2 \to 0$, $\P[\widehat{\T} \leq u] \to \Phi(u)$, where $\Phi(u)$ denotes the standard Gaussian cumulative distribution function. Importantly, as discussed by \cite{Calonico-Cattaneo-Titiunik_2014_ECMA} in the context of standard univariate RD designs, the usual confidence intervals 
\begin{align*}
    \widehat{\CI}(\alpha) 
    = \bigg[ \; \widehat{\tau}_{\mathtt{DIM}} - \mathfrak{c}_{\alpha} \cdot \sqrt{\widehat{\mathbb{V}}[\widehat{\tau}_{\mathtt{DIM}}]}
             \;\; , \;\;
             \widehat{\tau}_{\mathtt{DIM}} + \mathfrak{c}_{\alpha} \cdot \sqrt{\widehat{\mathbb{V}}[\widehat{\tau}_{\mathtt{DIM}}]}
             \; \bigg]
\end{align*}
with $\mathfrak{c}_{\alpha} = \Phi^{-1}(1 - \alpha/2)$, will not be valid whenever there is local misspecification. More precisely, the validity of the usual confidence intervals $\widehat{\CI}(\alpha)$ require the small bias condition $nh \mathsf{B}_n^2 \to 0$, which is not satisfied when the MSE-optimal bandwidth (or larger) is used. Therefore, the choice of bandwidth and regression specification determines whether uncertainty quantification is valid. 

If the MSE-optimal bandwidth is used, which in the context of pooling-based methods would depend on the order of the bias $\mathsf{B}_n$, then the small bias condition will not be satisfied, and thus the resulting inference procedures will over-reject the null hypothesis, sometimes substantially. Following \cite{Calonico-Cattaneo-Titiunik_2014_ECMA}, a simple and practical solution is to employ robust bias correction. The two-step procedure is simple and intuitive: first, the pooled treatment effect is estimated using a choice of polynomial approximation $p$ and its associated MSE-optimal bandwidth; second, confidence intervals and hypothesis tests are constructed by estimating the pooled treatment effect and variance estimator using the same bandwidth choice but with a polynomial of larger order $q$. In practice, the most common choice is $p=1$ for MSE-optimal treatment effect estimation, and $q=2$ for robust bias-corrected uncertainty quantification. This inference approach has several theoretical advantages \citep{Calonico-Cattaneo-Farrell_2018_JASA,Calonico-Cattaneo-Farrell_2022_Bernoulli}, and has been validated empirically \citep{Hyytinen-Tukiainen-etal2018_QE,DeMagalhaes-etal_2025_PA}. Standard RD software for estimation and inference in univariate score settings (\texttt{rdrobust}, \texttt{rdhte}) available at \url{https://rdpackages.github.io/} can be used directly, provided the regularity conditions are satisfied.

\section{Heterogeneity and Aggregation Along the Boundary}\label{sec: Distance-based and Location-Based Methods}

Our review of the literature indicates that most empirical researchers using BD designs employ pooling-based methods, thereby focusing on (density-weighted) average treatment effects along the entire (or a few segments of the) boundary $\B$. In most implementations, the estimation begins by calculating the distance to the nearest boundary point for every observation, keeping only observations for which this distance is less than a chosen bandwidth, and then continues by pooling all observations in a single local regression analysis, mimicking standard univariate RD methods.

Although the focus on a single scalar parameter is a convenient way of summarizing the information provided by the BD design, this approach fails to exploit its full richness. In this section, we discuss a more general approach for the analysis of BD designs that starts by studying the average treatment effect at each point on the boundary, and then uses this functional causal parameter as the building block for aggregation along the boundary, hence recovering other causal treatment effects. This section summarizes the recent results on identification, estimation, and inference discussed in \cite{Cattaneo-Titiunik-Yu_2025_BDD-Distance,Cattaneo-Titiunik-Yu_2025_BDD-Location}.

The starting point is to consider the Boundary Average Treatment Effect Curve (BATEC):
\begin{align*}
    \tau(\bx) = \E[ Y_i(1) - Y_i(0) | \bX_i=\bx],
    \qquad \bx\in\B,
\end{align*}
which was already implicitly introduced as part of the probability limit emerging from the pooling approaches. However, the functional causal parameter $\tau(\bx)$ is of independent interest because it captures the average treatment effects at each point on the boundary. Despite capturing potentially interesting causal evidence of heterogeneous treatment effects, only a handful of empirical papers have investigated the BATEC or variations thereof. Exceptions include \cite{Keele-etal_2017_AIE}, \cite{velez2019tuning}, and \cite{gonzalez2021cell}.

Another feature of the BATEC is that it provides a key building block for constructing other causal parameters of interest by aggregation along the boundary. For example, \cite{Cattaneo-Titiunik-Yu_2025_BDD-Location} discuss the following two parameters.
\begin{itemize}
    \item Weighted Boundary Average Treatment Effect (WBATE):
    \begin{align*}
        \tau_{\mathtt{WBATE}} = \frac{\int_{\B} \tau(\bx) w(\bx) \text{d}\Haus(\bx)}
                                     {\int_{\B} w(\bx) \text{d}\Haus(\bx)},
    \end{align*}
    where $w(\bx)$ denotes a user-chosen weighting scheme.
    
    \item Largest Boundary Average Treatment Effect (LBATE):
    \begin{align*}
        \tau_{\mathtt{LBATE}} = \sup_{\bx\in\B} \tau(\bx).
    \end{align*}
\end{itemize}

The WBATE parameter represents a weighted average of the (potentially heterogeneous) treatment effects $\tau(\bx)$ at each boundary point. In particular, setting $w(\bx)=f(\bx)$ recovers the BATE parameter $\tau$. In practice, other weighting schemes may also be of interest. See \cite{Reardon-Robinson_2012_JREE} and \cite{Wong-Steiner-Cook_2013_JEBS} for early methodological discussions, and \cite{Rischard-Branson-Miratrix-Bornn_2021_JASA} for other examples considered in the context of Bayesian methods for BD designs. The insightful discussion of \cite{Qiu-Stoye_2026_BookCh--Discussion} provides another application of the WBATE parameter. The LBATE parameter captures the ``best'' causal treatment effect along the assignment boundary, and thus can be useful for evaluating or developing targeted policies. See, for example, \cite{Andrews-Kitagawa-McCloskey_2024_QJE}.

\begin{figure}
    \centering
    \begin{subfigure}[b]{0.45\textwidth}
        \centering
        \includegraphics[width=\linewidth]{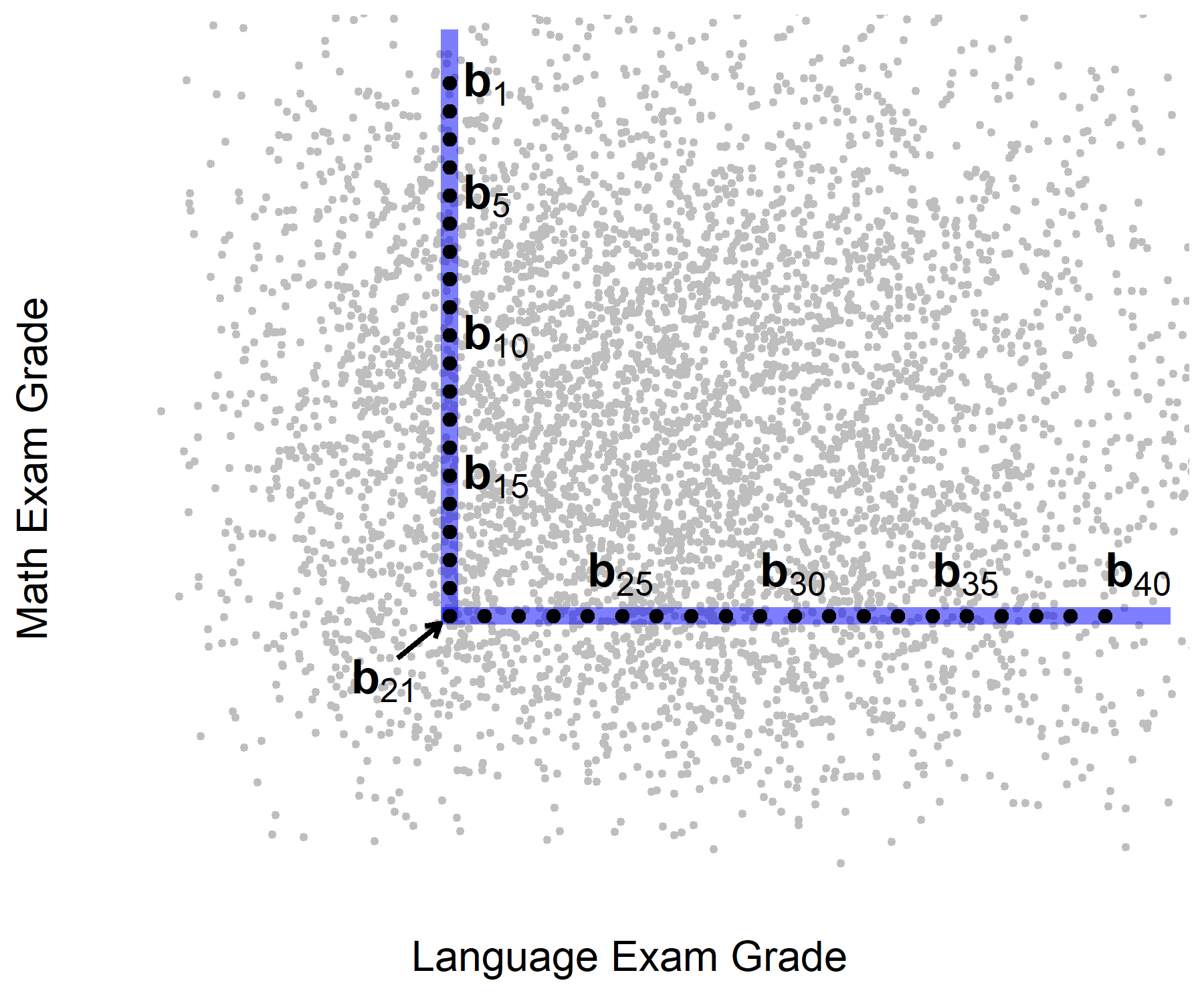}
        \caption{Boundary Discretization}
        \label{fig:HetAgg-BoundaryGrid}
    \end{subfigure}
    \hspace{.25in}
    \begin{subfigure}[b]{0.45\textwidth}
        \centering
        \includegraphics[width=\linewidth]{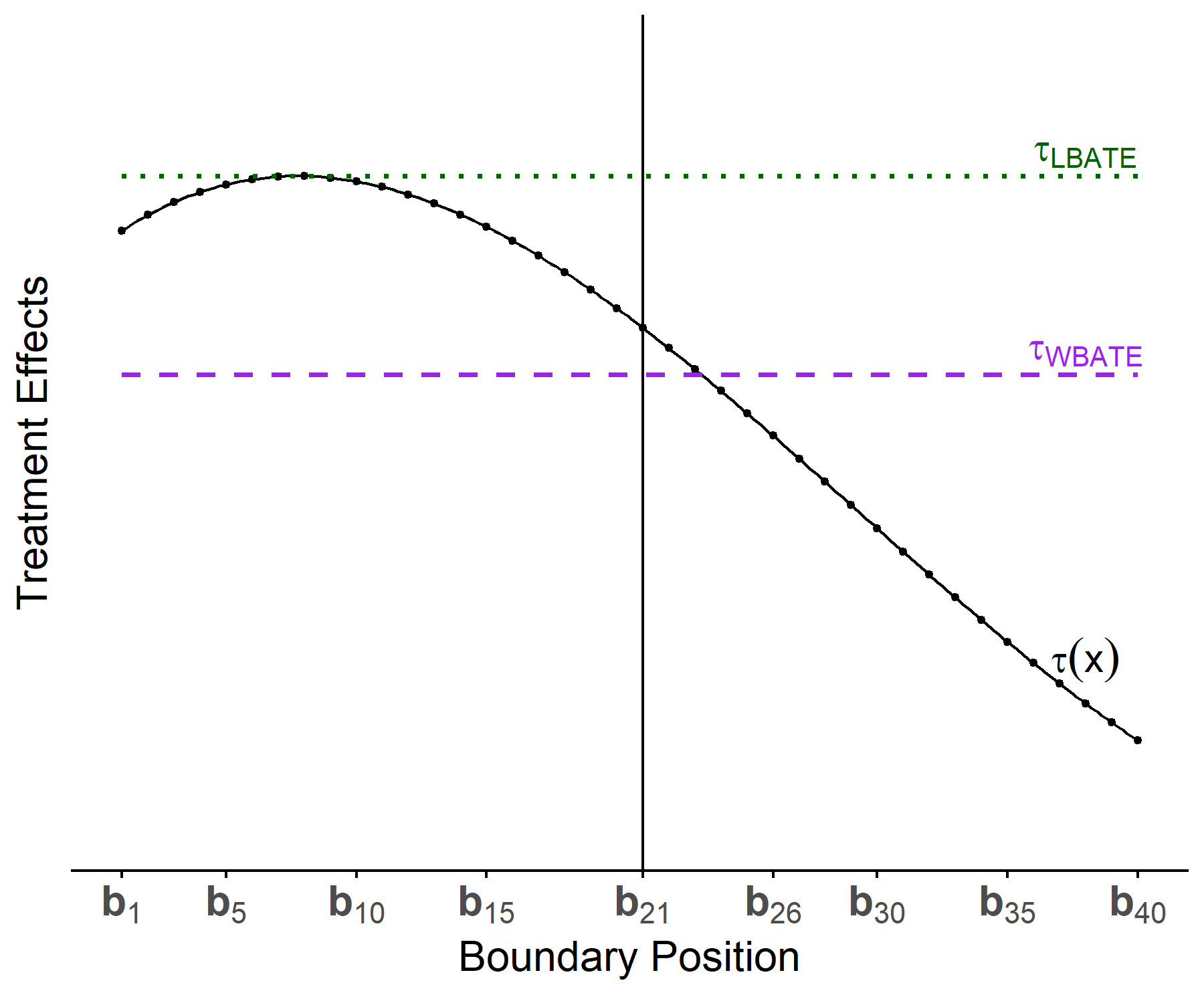}
        \caption{Causal Parameters}
        \label{fig:HetAgg-Parameters}
    \end{subfigure}    
    \caption{Heterogeneity and Aggregation along the Boundary}
    \label{fig:HetAgg}
\end{figure}

Figure \ref{fig:HetAgg} gives a graphical representation of the BATEC, WBATE, and LBATE in the context of the same stylized BD design used in Section \ref{sec: Pooling-Based Methods}. Because $\tau(\bx)$ is a function, estimation and inference proceed by discretizing the boundary to construct local regression estimates for each point $\bx\in\B$ on the grid of evaluation points chosen. Figure \ref{fig:HetAgg-BoundaryGrid} demonstrates the idea with $40$ evenly-spaced grid points, denoted by $\bb_1,\ldots,\bb_{40}$. As discussed in the upcoming subsections, distance-based and location-based methods employ these points to conduct estimation and inference, both pointwise and uniformly along the assignment boundary.

Figure \ref{fig:HetAgg-Parameters} illustrates graphically the three target causal parameters $(\tau(\bx):\bx\in\B)$, $\tau_{\mathtt{WBATE}}$, and $\tau_{\mathtt{LBATE}}$. The functional causal parameter $\tau(\bx)$ indicates the presence of heterogeneity along a region of the boundary, while $\tau_{\mathtt{WBATE}}$ and $\tau_{\mathtt{LBATE}}$ give two distinct notions of aggregation of those heterogeneous treatment effects. This stylized example is motivated by the SPP study by \cite{LondonoVelezRodriguezSanchez_2020_AEJ} introduced above, where $\bX_i= (X_{1i},X_{2i}) = (\text{\textit{SABER11}}_i,\text{\textit{SISBEN}}_i)$. Re-analyzing the SPP application, \cite{Cattaneo-Titiunik-Yu_2025_BDD-Location} found little evidence of heterogeneous treatment effects along the $X_{1i}$ dimension, but declining treatment effects along the $X_{2i}$ dimension. This empirical finding indicates that the average treatment effects along the boundary are roughly similar for all students with low academic performance ($\text{\textit{SABER11}}_i = 0$) and regardless of their wealth level ($\text{\textit{SISBEN}}_i \geq 0$), while heterogeneity in treatment effects is present for the wealthier eligible students ($\text{\textit{SISBEN}}_i = 0$) as their academic performance increases ($\text{\textit{SABER11}}_i \geq 0$).

The discussion so far has focused on causal parameters defined along the entire $\B$, but in some applications researchers have considered different segments of $\B$. Specification \eqref{eq: reg-Pooling-D-SFE-INTER-SEGHET}, and its motivation from the implementation used by \cite{LondonoVelezRodriguezSanchez_2020_AEJ} and others, gives one example. Employing the BATEC parameter as a building block, it is natural to define causal parameters as integrals or suprema of $\tau(\bx)$ over a specific region of $\B$. All the methods developed in \cite{Cattaneo-Titiunik-Yu_2025_BDD-Distance,Cattaneo-Titiunik-Yu_2025_BDD-Location} immediately apply to those cases, offering estimation and inference methods for segment-specific causal parameters based on the core BATEC parameter.

\subsection{Distance-Based Methods}

The distance $D_i=\d(\bX_i,\B)$ used by all pooling approaches is based on the distance of each observation $\bX_i$ to its nearest point $\bx\in\B$ on the boundary, denoted by $\d(\bX_i,\bx)$. Researchers sometimes first construct the signed distance to a pre-specified  point on the boundary
\begin{align*}
    D_i(\bx) = \big(\Indicator(\bX_i \in \A_1) - \Indicator(\bX_i \in \A_0)\big) \cdot \d(\bX_i, \bx),
    \qquad
    \bx \in \B,
\end{align*}
for each unit $i=1,\ldots,n$, and then use this distance to construct a tubular neighborhood covering $\B$ to implement pooling methods. One example of this approach is \citet{ehrlich2018persistent}. While the resulting tubular neighborhood will often be geometrically different from $\mathcal{T}(h) = \{\bx\in\X: \d(\bx,\B)\leq h\}$, all the ideas and results discussed in Section \ref{sec: Pooling-Based Methods} continue to apply.

However, the intermediate boundary-point-specific signed distance $(D_i(\bx):\bx\in\B)$ for each unit can also be used directly to estimate and conduct inference for the BATEC. The idea is to view the outcomes and the boundary-point-specific scalar distance variables, $(Y_1,D_1(\bx)),\ldots,(Y_n,D_n(\bx))$, as a one-dimensional RD design for each point $\bx\in\B$. For example, this approach was used by \cite{Keele-Titiunik_2015_PA} and \cite{velez2019tuning} to study the effect of TV exposure on voter turnout, using reception and media market boundaries that determine whether citizens are exposed to specific television programming. In these examples, each unit's score $\bX_i$ is a latitude-longitude pair that determines their place of residence, the outcome $Y_i$ is whether the individual turned out to vote, and the boundary $\B$ is the media market or reception boundary that separates the treated from the control region. In \cite{Keele-Titiunik_2015_PA}, the treatment is exposure to presidential candidate television ads during the presidential election, and in \cite{velez2019tuning} the treatment is exposure to Spanish-language television. In both cases, researchers first selected a grid $\bb_1,\bb_2,\ldots\bb_J$ on the boundary (e.g. Figure \ref{fig:HetAgg-BoundaryGrid}), and then used the local regression specification:
\begin{align}\label{eq: reg-Distance-INTER}
    \reg \quad Y_i \quad \mathtt{on} 
         \quad 1, \; T_i, \; \br_p(D_i(\bb_j)), \; T_i\cdot\br_p(D_i(\bb_j))
         \qquad [ \; \Indicator( |D_i(\bb_j)|\leq h ) \; ],
\end{align}
for each $j=1,\ldots, J$. The key difference is that estimation is conducted with localization for each evaluation point $\bb_j$ on the boundary, as opposed to localization to the entire boundary as in specifications \eqref{eq: reg-Pooling-DIM} through \eqref{eq: reg-Pooling-D-SFE-INTER-SEGHET}. Therefore, the estimated coefficient accompanying $T_i$ gives a point estimate of $\tau(\bb_j)$ for each $j=1,\ldots, J$.

\cite{Cattaneo-Titiunik-Yu_2025_BDD-Distance} study the large sample properties of the distance-based estimation procedure for BATEC obtained from specification \eqref{eq: reg-Distance-INTER}. They give necessary and/or sufficient conditions for identification, estimation, and inference in large samples, both pointwise and uniformly along the boundary. Their identification result requires minimal regularity of the boundary $\B$, the distance function $\d(\cdot)$, and the underlying data generating process, building on continuity-based identification in standard univariate RD designs \citep{Hahn-Todd-vanderKlaauw_2001_ECMA}. \cite{Cattaneo-Titiunik-Yu_2025_BDD-Distance} also show that the bias of the distance-based point estimator will depend on the smoothness of the assignment boundary, while the variance of the estimator will be of order $(nh^2)^{-1}$. Putting these two results together, it follows that bandwidth selection methods for standard univariate RD designs will either minimize the mean square error of the estimator or deliver undersmoothing, depending on the geometry of the assignment boundary and other assumptions. For inference, the authors establish asymptotic validity of confidence intervals and confidence bands, provided the bias of the distance-based point estimator is small enough (which, in turn, crucially depends on the geometry of $\B$).

\subsection{Location-Based Methods}

While the distance-based estimator constructed using specification \eqref{eq: reg-Distance-INTER} provides a natural bridge between the pooling methods and heterogeneity analysis, \cite{Cattaneo-Titiunik-Yu_2025_BDD-Distance} demonstrate that this approach would require a smooth assignment boundary for higher-order bias reduction. Otherwise, the distance-based approach will generate a consistent estimator with possibly a high bias, and therefore require a small bandwidth for estimation and inference validity. A solution to this problem is to use local \textit{bivariate} regression methods.

\cite{Cattaneo-Titiunik-Yu_2025_BDD-Location} study the properties of local regression treatment effect estimators directly employing the location information encoded in $\bX_i$ for each unit. The authors develop pointwise and uniform estimation and inference methods for both the BATEC and transformations thereof, such as the WBATE and LBATE. Their local regression specification takes the form:
\begin{align}\label{eq: reg-Location-INTER}
    \reg \quad Y_i \quad \mathtt{on} 
         \quad 1, \; T_i, \; \br_p(\bX_i - \bb_j), \; T_i\cdot\br_p(\bX_i - \bb_j)
         \qquad [ \; K\big(\tfrac{\bX_i - \bb_j}{h}\big) \; ],
\end{align}
where $\bb_1,\ldots,\bb_J$ is the grid of points on the boundary (e.g., Figure \ref{fig:HetAgg-BoundaryGrid}). This local polynomial specification crucially depends on $\bX_i$ directly via the $p$th order polynomial expansion $\br_p(\bX_i - \bx)$. The weights are determined by the bivariate kernel function $K(u_1,u_2)$, and the localization to each point $\bb_j$ continues to be is controlled by the bandwidth $h$. In particular, $K\big(\frac{\bX_i - \bb_j}{h}\big) = \Indicator( \|\bX_i - \bb_j\| \leq h) = \Indicator( | D_i(\bb_j)| \leq h)$ is a valid choice, demonstrating an interesting connection with distance-based methods as implemented via specification \eqref{eq: reg-Distance-INTER}. The estimated coefficient accompanying $T_i$, denoted by $\widehat{\tau}(\bb_j)$, gives the location-based point estimate of $\tau(\bb_j)$, for each $j=1,\ldots,J$.

In the location-based setting, identification of $\tau(\bx)$ for $\bx\in\B$ follows directly from a generalization of continuity-based identification results in the standard univariate RD design \citep[see][]{Hahn-Todd-vanderKlaauw_2001_ECMA,Papay-Willett-Murnane_2011_JoE,Reardon-Robinson_2012_JREE,Wong-Steiner-Cook_2013_JEBS,Keele-Titiunik_2015_PA,Cattaneo-Keele-Titiunik-VazquezBare_2016_JOP}. Under minimal regularity conditions on the boundary and data generating process, \cite{Cattaneo-Titiunik-Yu_2025_BDD-Location} establishes (pointwise and) uniform convergence rates of the form $\sup_{\bx\in\B}|\widehat{\tau}(\bx)-\tau(\bx)|^2 = O_\P(\tfrac{\log(n)}{nh^2} + h^{2(p+1)})$, demonstrating that the location-based estimator does not require stringent smoothness restrictions on the boundary $\B$ to achieve higher order debiasing. Furthermore, the authors establish pointwise and integrated mean square error expansions for the estimator, and develop MSE-optimal bandwidth selection methods. By combining these results, they develop optimal point estimation of the BATEC, both pointwise and uniform, based on specification \eqref{eq: reg-Location-INTER}.

For uncertainty quantification, \cite{Cattaneo-Titiunik-Yu_2025_BDD-Location} derive confidence intervals and confidence bands of the form:
\begin{align*}
    \widehat{\CI}(\bx,\alpha) 
    = \bigg[ \; \widehat{\tau}(\bx) - \mathfrak{c}_{\alpha} \cdot \sqrt{\widehat{\mathbb{V}}[\widehat{\tau}(\bx)]}
             \;\; , \;\;
             \widehat{\tau}(\bx) + \mathfrak{c}_{\alpha} \cdot \sqrt{\widehat{\mathbb{V}}[\widehat{\tau}(\bx)]}
             \; \bigg],
    \qquad \bx\in\B,
\end{align*}
where $\widehat{\mathbb{V}}[\widehat{\tau}(\bx)]$ denotes any of the usual variance estimators from (weighted) least squares regression methods, and $\mathfrak{c}_{\alpha}$ is the quantile used depending on inferential goal. For confidence intervals, the usual Gaussian quantile is asymptotically valid (i.e., $\mathfrak{c}_{\alpha} = \Phi^{-1}(1 - \alpha/2)$), while for confidence bands, a new (larger) quantile needs to be constructed to capture the joint uncertainty features of the estimators $\widehat{\tau}(\bx)$ for different values along the boundary $\B$. As in the case of pooling-based and distance-based methods, a key condition needed for validity of these confidence intervals and bands is the small (misspecification) bias property. \cite{Cattaneo-Titiunik-Yu_2025_BDD-Location} address this issue by relying on robust bias correction \citep{Calonico-Cattaneo-Titiunik_2014_ECMA,Calonico-Cattaneo-Farrell_2022_Bernoulli}, a construction that proceeds in two steps.
\begin{itemize}
    \item \textit{Step 1}. For a choice $\mathsf{P}$ (usually $\mathsf{P}=1$), the point estimator $\widehat{\tau}(\bx)$ is computed using specification \eqref{eq: reg-Location-INTER} with $p=\mathsf{P}$ and its corresponding MSE-optimal bandwidth. As a result, $\widehat{\tau}(\bx)$ is an asymptotically MSE-optimal point estimator of $\tau(\bx)$.
    
    \item \textit{Step 2}. For a choice $\mathsf{Q}>\mathsf{P}$ (usually $\mathsf{Q}=2$), the point estimator $\widehat{\tau}(\bx)$ and variance estimator $\widehat{\mathbb{V}}[\widehat{\tau}(\bx)]$ are computed using specification \eqref{eq: reg-Location-INTER} with $p=\mathsf{Q}$ and the same bandwidth used in \textit{Step 1}. These two quantities are used to construct the confidence interval estimator $\widehat{\CI}(\bx,\alpha)$. As a result, inference procedures are robust bias-corrected, and hence asymptotically valid.
\end{itemize}
\cite{Cattaneo-Titiunik-Yu_2025_BDD-Location} give further technical and computational details, which we omit to conserve space.

Finally, $\widehat{\tau}(\bx)$ can also be used to estimate and conduct inference for WBATE and LBATE via plug-in methods. \cite{Cattaneo-Titiunik-Yu_2025_BDD-Location} provide sufficient conditions and study the validity of such methods. In particular, the WBATE corresponds to an integral over a submanifold of the nonparametric estimator $\widehat{\tau}(\bx)$. See \cite{chen2025semiparametric} for related theoretical results when employing series approximations instead of local polynomial regression as commonly done in RD settings.

\begin{figure}
    \centering
    \includegraphics[scale=0.65]{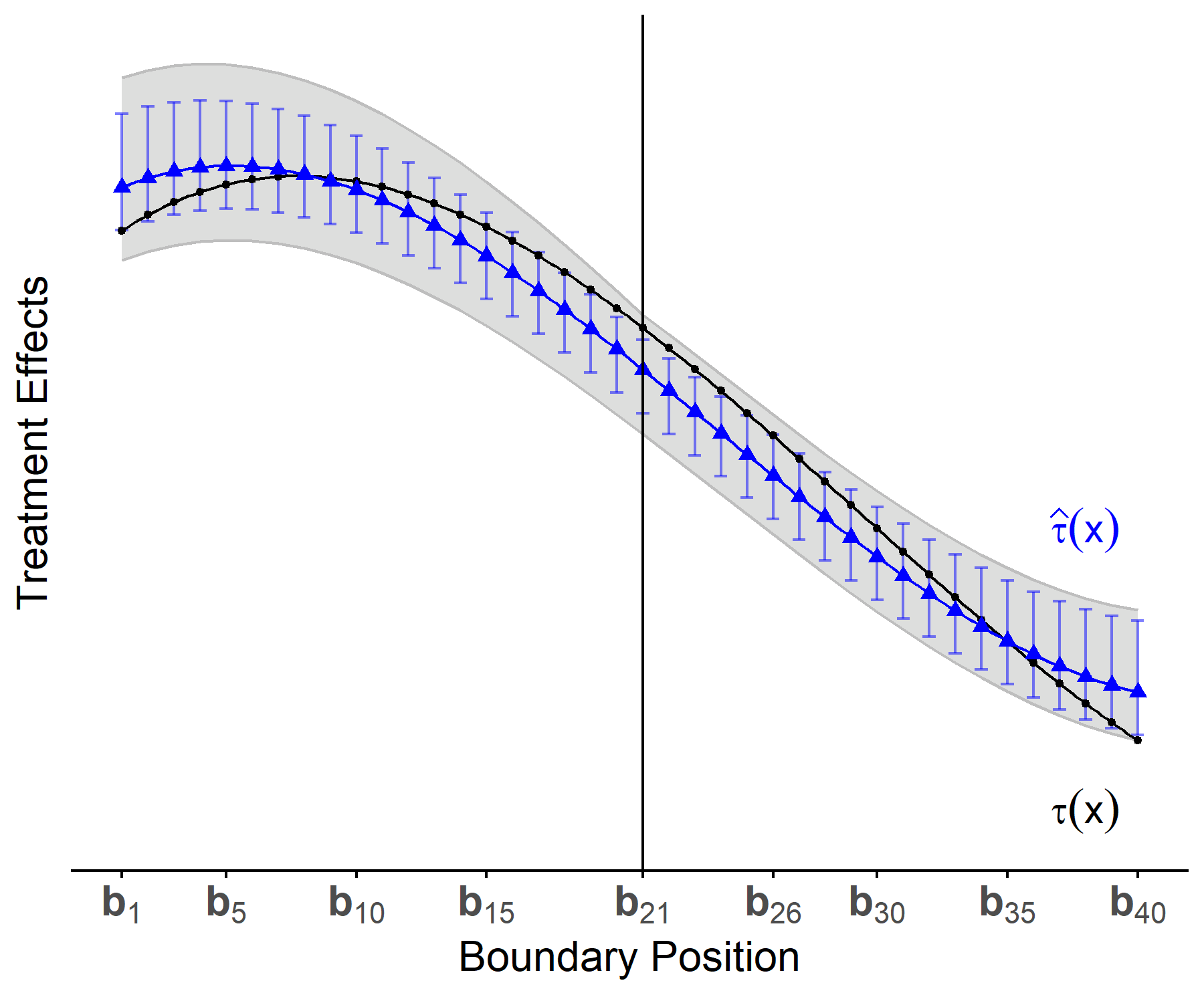}
    \caption{Estimation and Inference for BATEC}
    \label{fig:CIandCB}
\end{figure}

The general-purpose \texttt{R} package \texttt{rd2d} implements both distance-based and location-based methods (\url{https://rdpackages.github.io/}), see \cite{Cattaneo-Titiunik-Yu_2025_rd2d} for details. Figure \ref{fig:CIandCB} gives a stylized example of the type of results that can be obtained via distance-based and location-based methods, building on the same setup as in Figure \ref{fig:BDillustration}. These figures are motivated by empirical work using the SPP application \citep{LondonoVelezRodriguezSanchez_2020_AEJ}.

\section{Recommendations for Practice} \label{sec: Recommendations for Practice}

Our review of the empirical literature employing BD designs showed that the overwhelming majority of studies focuses exclusively on identification, estimation, and inference for the BATE using pooling-based methods (Section \ref{sec: Pooling-Based Methods}).
This scalar parameter captures a density-weighted average of potentially heterogeneous treatment effects along the assignment boundary. While modern empirical work has recognized the importance of localization and flexible local parametrization, the current literature often exhibits three key limitations: (i) local regression specifications frequently omit the interaction between the treatment assignment variable $T_i$ and the flexible regression terms $\br_p(D_i)$ or $\br_p(\bX_i)$, risking significant misspecification bias; (ii) the bandwidth $h$ is often chosen in an ad-hoc manner; and (iii) the role of misspecification bias from a large bandwidth, such as an MSE-optimal choice, is often ignored, which can lead to over-rejection of the null hypothesis. \cite{Cattaneo-Titiunik-Yu_2025_BDD-Pooling} provide a formal study of pooling-based methods that explicitly addresses these limitations. Moving forward, we recommend that empirical researchers implementing pooling-based methods use specifications \eqref{eq: reg-Pooling-D-SFE-INTER} or \eqref{eq: reg-Pooling-X-SFE-INTER}, select an MSE-optimal bandwidth, and conduct inference using robust bias-corrected methods.

As noted by \cite{Cattaneo-Titiunik-Yu_2025_BDD-Distance,Cattaneo-Titiunik-Yu_2025_BDD-Location}, a more general approach to analyzing and interpreting BD designs is to first focus on the BATEC parameter, and then consider other aggregate causal treatment effects as transformations thereof. As summarized in Section \ref{sec: Distance-based and Location-Based Methods}, distance-based and location-based methods capture the full richness of the BD design by estimating average treatment effects at each point along the assignment boundar, and also recover the BATE parameter as a special case. These methods can be implemented using the general-purpose \texttt{R} package \texttt{rd2d} available at \url{https://rdpackages.github.io/}, and discussed in \cite{Cattaneo-Titiunik-Yu_2025_rd2d}. In particular, location-based methods are the most robust and general approach; we therefore recommend them for future empirical work analyzing BD designs.

Finally, our discussion and literature review have focused on BD designs where the running variable is explicitly bivariate, such as geographic location or test scores in two exams. A more general view of BD designs also includes univariate RD designs with multiple cutoffs. As discussed by \cite{CattaneoKeeleTitiunikVaquezBare2016-JOP} and \cite{Cattaneo-Idrobo-Titiunik_2024_CUP}, a univariate RD design with multiple cutoffs can be recast as a BD design where one dimension of the bivariate score is the original univariate score and the other is the cutoff variable, and the boundary is the set where these dimensions are equal---for an empirical example where multiple cutoffs induce a known boundary see the seminal paper by \cite{AngristLavy1999-QJE}. In this sense, with appropriate modifications, our discussion of BD design methodology also applies to RD designs with multiple cutoffs. 

\clearpage
\singlespacing
\renewcommand*{\arraystretch}{1.5}
\setlength{\tabcolsep}{4pt} 

\begin{landscape}

\begin{footnotesize} 

\begin{xltabular}{\textwidth}{
    >{\raggedright\arraybackslash}p{3cm}              
    >{\centering\arraybackslash}p{3cm}                
    >{\centering\arraybackslash}p{3cm}                
    >{\centering\arraybackslash}p{3cm}                
    >{\centering\arraybackslash}p{3cm}                
    >{\centering\arraybackslash}p{3cm}                
    >{\centering\arraybackslash}p{2cm}                
    >{\centering\arraybackslash}p{1cm}                
    }

    \caption{Selected Empirical Articles Using The Boundary Discontinuity Design}\label{tab:LitReview}\\
    
    \toprule
    Article & $Y_i$ & $T_i$ & $\bX_i$ & Distance & Specifications & Bandwidth & Heter\\
    \midrule
    \endfirsthead

    \multicolumn{8}{c}{{\tablename\ \thetable{} -- Continued from previous page}}\\
    \toprule
    Article & $Y_i$ & $T_i$ & $\bX_i$ & Distance & Specifications & Bandwidth & Heter\\
    \midrule
    \endhead

    \midrule
    \multicolumn{8}{r}{{Continued on next page...}} \\
    \endfoot
   
    \bottomrule
   \multicolumn{8}{p{23cm}}{\footnotesize {Note: MSE denotes optimal mean squared error bandwidth choice. Binding score refers to the approach of constructing a single univariate score as the minimum of all scores that determine treatment assignment, known as the binding score, and then conducting a univariate RD analysis using the binding score  \citep[see, e.g.,][]{Reardon-Robinson_2012_JREE}. A specification number with the symbol $^\star$ indicates it has been modified in a some way; for details, we refer readers to the article. GPA denotes grade point average.}}
    \endlastfoot
    
\citet{Card-Krueger_1994_AER} & Employment & Minimum wage increase & Location of towns/counties & NA & (\ref{eq: reg-Pooling-DIM}), (\ref{eq: reg-Pooling-DIM-SFE}) & NA & No \\ 
  \citet{holmes1998effect} & Manufacturing activity & Right-to-work policies & Lat, Lon & Euclidean & (\ref{eq: reg-Pooling-DIM}), (\ref{eq: reg-Pooling-DIM-SFE}), (\ref{eq: reg-Pooling-D-SFE-INTER}) & Manual & Yes \\ 
  \citet{black1999better} & House prices & School quality & Lat, Lon & Euclidean & (\ref{eq: reg-Pooling-DIM}), (\ref{eq: reg-Pooling-DIM-SFE}) & Manual & No \\ 
  \citet{kane2003school} & House prices & School quality & Lat, Lon & Euclidean & (\ref{eq: reg-Pooling-DIM}), (\ref{eq: reg-Pooling-DIM-SFE}) & Manual & No \\ 
  \citet{kane2006school} & Housing prices & School quality & Lat, Lon & Euclidean & (\ref{eq: reg-Pooling-DIM}), (\ref{eq: reg-Pooling-DIM-SFE})$^\star$ & Manual & No \\ 
  \citet{bayer2007unified} & House prices & School quality & Lat, Lon & Euclidean & (\ref{eq: reg-Pooling-DIM}), (\ref{eq: reg-Pooling-DIM-SFE}) & Manual & No \\ 
  \citet{lalive2008extended} & Unemployment duration & Unemployment benefits & Lat, Lon & Driving time & (\ref{eq: reg-Pooling-D-SFE-INTER}) & Manual & No \\ 
  \citet{dell2010persistent} & Consumption, childhood stunting & Forced mining labor & Lat, Lon & Euclidean & (\ref{eq: reg-Pooling-X-SFE}) & MSE & No \\ 
  \citet{ou2010leave} & High school dropout & Failure of high school exit exam & Mathematics score, Language score & Euclidean & (\ref{eq: reg-Pooling-D-SFE-INTER-SEGHET}) & Manual & No \\ 
  \cite{dube2010minimum} & Employment & Increase in minimum wage & Lat, Lon & County adjacency & (\ref{eq: reg-Pooling-DIM}), (\ref{eq: reg-Pooling-DIM-SFE}) & Manual & No \\ 
  \citet{grout2011land} & Property values & Land-use regulations & Lat, Lon & Euclidean & (\ref{eq: reg-Pooling-D-SFE-INTER-SEGHET}) & Manual & Yes \\ 
  \citet{robinson2011evaluating} & English proficiency & English proficiency reclassification & Scores on five language tests & Euclidean & Binding score & MSE & No \\ 
  \cite{eugster2011demand} & Demand for social insurance & Culture & Lat, Lon & Driving kilometers & (\ref{eq: reg-Pooling-D-SFE-INTER}) & Manual & No \\ 
  \citet{hinnerich2014democracy} & Political regimes & Income redistribution & Population in two separate years & Euclidean & (\ref{eq: reg-Pooling-D-SFE-INTER-SEGHET}) & MSE & No \\ 
  \citet{ferwerda2014political} & Resistance activity & Devolution of governing authority & Lat, Lon & Euclidean & (\ref{eq: reg-Pooling-D-SFE-INTER}) & MSE, Manual & No \\ 
  \citet{snider2015barriers} & Airline industry outcomes & Access to airport facilities granted to new airlines & Share of passengers in top two carriers in origin and destination city &  & (\ref{eq: reg-Pooling-DIM}), (\ref{eq: reg-Location-INTER}) & MSE, Manual & Yes \\ 
  \citet{barone2015telecracy} & Vote shares & Switch from analog to digital TV & Lat, Lon & Euclidean & (\ref{eq: reg-Pooling-D-SFE}) & Manual & No \\ 
  \citet{michalopoulos2016long} & Civil conflict & Ethnic partitioning & Lat, Lon & Euclidean & (\ref{eq: reg-Pooling-DIM}), (\ref{eq: reg-Pooling-DIM-SFE}) & Manual & No \\ 
  \citet{egger2015impact} & Foreign investment & Controlled foreign company rules & Tax rate, Income measures & NA & (\ref{eq: reg-Pooling-X-SFE-INTER}) $L=1$ & MSE & No \\ 
  \cite{Keele-Titiunik_2015_PA} & Voter turnout & Political TV advertisement & Lat, Lon & Chordal & (\ref{eq: reg-Distance-INTER}) & MSE & Yes \\ 
  \citet{macdonald2016effect} & Crime & Policing & Lat, Lon & Euclidean & (\ref{eq: reg-Pooling-D-SFE-INTER}) & MSE & No \\ 
  \citet{evans2017smart} & STEM major & Financial incentive & GPA,  Family contribution score & Euclidean & (\ref{eq: reg-Pooling-D-SFE-INTER-SEGHET}) & MSE & No \\ 
  \cite{eugster2017culture} & Unemployment & Culture & Lat, Lon & Road distance & (\ref{eq: reg-Pooling-D-SFE-INTER}) & Manual & No \\ 
  \citet{kumar2018restrictions} & Mortgage default & Limits on home equity borrowing & Lat, Lon & Euclidean & (\ref{eq: reg-Pooling-D}), (\ref{eq: reg-Pooling-X-SFE}) & Manual & No \\ 
  \citet{ehrlich2018persistent} & Economic density & Subsidies & Lat, Lon & Euclidean & (\ref{eq: reg-Pooling-D})+$\br_p(\bX_i)$ & MSE, Manual & No \\ 
  \citet{dell2018historical} & Economic outcomes & Dai Viet administrative institutions & Lat, Lon & Euclidean & (\ref{eq: reg-Pooling-X-SFE}), (\ref{eq: reg-Pooling-D-SFE}), (\ref{eq: reg-Pooling-D-SFE})+$\br_p(\bX_i)$ & Manual & No \\ 
  \citet{dell2018nation} & Public goods, political attitudes & War strategies & Lat, Lon & Euclidean & (\ref{eq: reg-Pooling-X-SFE}) & Manual & No \\ 
  \citet{clinton2018politics} & Voter registration and turnout & Medicaid expansion & Lat, Lon & Euclidean & (\ref{eq: reg-Pooling-D-SFE-INTER}) & Manual & No \\ 
  \citet{spenkuch2018political} & Voter turnout & Television advertisement & Lat, Lon & Euclidean & (\ref{eq: reg-Pooling-DIM-SFE}), (\ref{eq: reg-Pooling-D-SFE-INTER}) & MSE, Manual & No \\ 
  \citet{de2018agents} & Party vote shares & Political alignment & Lat, Lon & Euclidean & (\ref{eq: reg-Pooling-X-SFE}) & Manual & No \\ 
  \citet{eugster2019culture} & Tax competition & Culture & Lat, Lon & Road & (\ref{eq: reg-Pooling-DIM}) in DID & Manual & No \\ 
  \citet{giuntella2019sunset} & Sleep duration, Health outcomes & Timing of natural light & Lat, Lon & Euclidean & (\ref{eq: reg-Pooling-D-SFE-INTER}) & MSE, Manual & No \\ 
  \citet{johnson2019effects} & High School graduation & English learner classification & Scores on two language tests & Euclidean & Binding score & MSE & No \\ 
  \citet{velez2019tuning} & Voter turnout & Spanish-language television station & Lat, Lon & Euclidean & (\ref{eq: reg-Pooling-DIM}), (\ref{eq: reg-Pooling-DIM-SFE}), (\ref{eq: reg-Distance-INTER}) & MSE, Manual & Yes \\ 
  \citet{dupraz2019french} & Schooling & Colonial legacies & Lat, Lon & Euclidean & (\ref{eq: reg-Pooling-D-SFE-INTER}), (\ref{eq: reg-Pooling-X-SFE}) & MSE & No \\ 
  \citet{moscona2020segmentary} & Conflict & Segmentary lineage organization & Lat, Lon & Euclidean & (\ref{eq: reg-Pooling-D-SFE-INTER}) & Manual & No \\ 
  \citet{ambrus2020loss} & Property value & Cholera epidemic & Lat, Lon & Walking & (\ref{eq: reg-Pooling-D-SFE-INTER}) $L=1$ & MSE, Manual & No \\ 
  \citet{dell2020development} & Schools, education levels & Sugar cultivation for Dutch cultivation system & Lat, Lon & Euclidean & (\ref{eq: reg-Pooling-X-SFE}) & Manual & No \\ 
  \citet{he2020watering} & Pollution & Water quality monitoring & Lat, Lon & Euclidean & (\ref{eq: reg-Pooling-D-SFE-INTER}) $L=1$ & MSE & No or NA \\ 
  \citet{ito2020willingness} & Willingness to pay for clean air &  & Lat, Lon & Euclidean, road & (\ref{eq: reg-Pooling-D-SFE-INTER}) & MSE, Manual & No \\ 
  \citet{wuepper2020countriesErosion} & Soil erosion & Country borders & Lat, Lon & Euclidean & (\ref{eq: reg-Pooling-D-SFE-INTER}) & MSE & No \\ 
  \citet{albertus2020land} & Conflict & Land reform & Lat, Lon & Euclidean & (\ref{eq: reg-Pooling-D-SFE-INTER}) & MSE, Manual & No \\ 
  \citet{wuepper2020countriesPollution} & Crop yield gaps, nitrogen pollution & Country borders & Lat, Lon & Euclidean & (\ref{eq: reg-Pooling-D-SFE-INTER}) & MSE & No \\ 
  \citet{schafer2020time} & Voter turnout & Time zones & Lat, Lon & Chordal & (\ref{eq: reg-Pooling-D-SFE-INTER}) & MSE, Manual & No \\ 
  \citet{letsa2020mechanisms} & Attitudes towards local power & Colonial legacies & Lat, Lon & Euclidean & (\ref{eq: reg-Pooling-X-SFE}) & Manual & No \\ 
  \cite{LondonoVelezRodriguezSanchez_2020_AEJ} & College enrollment & Financial subsidy & High school exit exam score, wealth index & Euclidean & (\ref{eq: reg-Pooling-D-SFE-INTER-SEGHET}) & MSE & No \\ 
  \citet{gonzalez2021cell} & Election fraud & Cell phone coverage & Lat, Lon & Euclidean & (\ref{eq: reg-Pooling-D-SFE-INTER}), (\ref{eq: reg-Pooling-X-SFE-INTER}) & MSE & Yes \\ 
  \citet{ayres2021environmental} & Land value & Adjudication of groundwater rights & Lat, Lon & Euclidean & (\ref{eq: reg-Pooling-D-SFE-INTER}) $L=1$ & MSE & No \\ 
  \citet{laliberte2021long} & School quality & Educational achievement & Lat, Lon & Euclidean & (\ref{eq: reg-Pooling-D-SFE-INTER}) & MSE & No \\ 
  \citet{aaronson2021effects} & Characteristics of urban neighborhoods & Red lining maps & Lat, Lon & Euclidean & (\ref{eq: reg-Pooling-DIM-SFE}) in DID & Manual & No \\ 
  \citet{appau2021long} & Agricultural productivity & Bombing intensity & Lat, Lon & Absolute difference between centroid and 17th parallel north latitude & (\ref{eq: reg-Pooling-D}) (first stage) & Manual & No \\ 
  \citet{lowes2021concessions} & Education, health, wealth & Concession to extract rubber & Lat, Lon & Euclidean & (\ref{eq: reg-Pooling-D-SFE-INTER}) & MSE, Manual & No \\ 
  \citet{sides2022effect} & Election outcomes & Television advertisement & Lat, Lon & Euclidean & (\ref{eq: reg-Pooling-DIM-SFE}) & Manual & No \\ 
  \citet{zheng2022valuation} & House prices & Charter school entry & Lat, Lon & Euclidean & Event study in sample near boundary & Manual & No \\ 
  \citet{dehdari2022origins} & Regional identity & Actions of nation states & Lat, Lon & Euclidean & (\ref{eq: reg-Pooling-D-SFE-INTER}) & MSE, Manual & No \\ 
  \citet{mendez2022multinationals} & Living standards & Land concession to multinational firm & Lat, Lon & Euclidean & (\ref{eq: reg-Pooling-X-SFE}) & Manual & No \\ 
  \citet{jones2022factor} & Profits, irrigation efficiency & Irrigation & Lat, Lon & Euclidean & (\ref{eq: reg-Pooling-D-SFE-INTER}) & Manual & No \\ 
  \citet{henn2023complements} & State capacity & Remoteness from administrative headquarters & Lat, Lon & Euclidean & (\ref{eq: reg-Pooling-D-SFE-INTER})$^\star$ & Manual & No \\ 
  \citet{jones2022effects} & Enrollment, graduation & Tuition scholarship & High school GPA, SAT/ACT score & NA & (\ref{eq: reg-Pooling-X-SFE-INTER}) & Manual & No \\ 
  \citet{dai2022effects} & Lifestyle outcomes & Hypertension diagnosis & Systolic and Diastolic blood pressure & Euclidean & (\ref{eq: reg-Pooling-D-SFE-INTER-SEGHET}) & MSE & No \\ 
  \citet{salti2022impact} & Economic outcomes & Cash assistance & Two scores in means testing formula & NA & (\ref{eq: reg-Pooling-D-SFE-INTER-SEGHET})$^\star$ & MSE & No \\ 
  \citet{castro2022effect} & Teacher retention, student learning & Recruitment bonus & Population, distance to capital & Euclidean & (\ref{eq: reg-Pooling-D-SFE-INTER-SEGHET}) & MSE & No \\ 
  \citet{moussa2022impact} & Children health outcomes & Cash assistance & Two scores in means testing formula & NA & (\ref{eq: reg-Pooling-D-SFE-INTER-SEGHET})$^\star$ & MSE & No \\ 
  \citet{mangonnet2022playing} & Protected area designation & Political alignment & Lat, Lon & Euclidean & (\ref{eq: reg-Pooling-DIM-SFE}) & Manual & No \\ 
  \citet{baragwanath2023collective} & Forest growth & Property rights & Lat, Lon & Euclidean & (\ref{eq: reg-Pooling-D-SFE-INTER}) & MSE & No \\ 
  \citet{larsen2024long} & Educational attainment & Grade retention & Mathematics and English Language Arts test scores & Euclidean & (\ref{eq: reg-Pooling-D-SFE-INTER-SEGHET}) & MSE & No \\ 
  \citet{woller2023cost} & Vote buying & Cost of voting & Lat, Lon & Euclidean & (\ref{eq: reg-Pooling-D}) & Manual & No \\ 
  \citet{paulsen2023foundations} & Voter turnout, economic outcomes & State funding for common schools & Lat, Lon & Euclidean & (\ref{eq: reg-Pooling-X-SFE}) & Manual & No \\ 
  \citet{prillaman2023strength} & Political participation & Participation in  women-only credit groups & Lat, Lon & Euclidean & (\ref{eq: reg-Pooling-DIM-SFE}), (\ref{eq: reg-Pooling-X-SFE}) & Manual & No \\ 
  \citet{mcalexander2023international} & Rebel actions & UN partition line of Palestine & Lat, Lon & Euclidean & (\ref{eq: reg-Pooling-D-SFE-INTER}) $L=1$ & MSE & No \\ 
  \citet{murphy2023dual} & Special education placement & English learner designation & Various scores of language assessment & NA & Binding score & MSE & No \\ 
  \citet{bjerre2024attendance} & School enrollment & School change & Lat, Lon & Euclidean & (\ref{eq: reg-Pooling-DIM-SFE}) in DID & Manual & No \\ 
  \citet{kampfen2024heterogeneous} & Blood pressure & Blood pressure diagnosis & Systolic and Diastolic blood pressure & NA & Binding score & MSE & No \\ 
  \citet{wuepper2024public} & Deforestation & Country borders & Lat, Lon & Euclidean & (\ref{eq: reg-Pooling-D-SFE-INTER}) & MSE & No \\ 
  \citet{cox2024bound} & Voter mobilization & Social networks & Lat, Lon & Fastest driving & (\ref{eq: reg-Pooling-DIM})$^\star$ & NA & No \\ 
  \citet{doucette2024parliamentary} & Urbanization, commercialization & Duchy of Württemberg & Lat, Lon & Euclidean & (\ref{eq: reg-Pooling-D})+$\br_p(\bX_i)$ & MSE, Manual & No \\ 
  \citet{doucette2024pre} & Party support & Inclusive
institutions & Lat, Lon & Euclidean & (\ref{eq: reg-Pooling-D})+$\br_p(\bX_i)$ & Manual & No \\ 
  \citet{grasse2024state} & Poverty, economic development & Mass repression & Lat, Lon & Euclidean & (\ref{eq: reg-Pooling-D-SFE}) & Manual & No \\ 
  \cite{jardim2024local} & Wages, employment & Increase in minimum wage & Lat, Lon & Driving time & (\ref{eq: reg-Pooling-DIM})$^\star$, (\ref{eq: reg-Pooling-DIM-SFE})$^\star$ & Manual & No \\ 
  \citet{ring2025wealth} & Savings & Wealth tax assessment & Lat, Lon & Euclidean & (\ref{eq: reg-Pooling-D-SFE-INTER}) & Manual & No \\ 
  \citet{yamagishi2025persistent} & Land price & Buraku neighborhoods & Lat, Lon & Euclidean & (\ref{eq: reg-Pooling-DIM}), (\ref{eq: reg-Pooling-D-SFE-INTER}) $L=1$ & MSE, Manual & No \\ 
  \citet{loumeau2025regional} & Commuting flows, residential decisions & Departmental borders & Lat, Lon & Euclidean & (\ref{eq: reg-Pooling-D-SFE}) & MSE & No \\ 
  \citet{boix2025political} & Jewish National identity & Political emancipation & Lat, Lon & Euclidean & (\ref{eq: reg-Pooling-D-SFE-INTER}) & MSE & No \\ 
  \citet{muller2025building} & Share of predominant ethnic group & Administrative borders & Lat, Lon & Euclidean & (\ref{eq: reg-Pooling-D-SFE-INTER}) & Manual & No \\
    
\end{xltabular}

\end{footnotesize}

\end{landscape}


\clearpage
\small
\onehalfspacing

\bibliography{Vol1_ch2_bib,Vol1_ch2_bib-empapp}
\bibliographystyle{ecta}

\end{document}